\documentclass[aps,nofootinbib,preprint,tightenlines,showpacs]{revtex4}%
\usepackage{amsfonts}
\usepackage{amsmath}
\usepackage{amssymb}
\usepackage{graphicx}%
\newcommand{\beq}{\begin{eqnarray}}
\newcommand{\eeq}{\end{eqnarray}}
\newcommand{\bqa}{\begin{eqnarray}}
\newcommand{\eqa}{\end{eqnarray}}

\def\mqo2{{\!\!\!}}

\begin{document}
\preprint{HISKP-TH-10-08}
\title{Causality and the effective range expansion}
\author{H.-W. Hammer$^{a}$ and Dean Lee$^{b,a}$}
\affiliation{$^{a}$Helmholtz-Institut f\"{u}r Strahlen- und Kernphysik (Theorie) and Bethe
Center for Theoretical Physics, Universit\"{a}t Bonn, D-53115 Bonn,
Germany\linebreak\linebreak$^{b}$Department of Physics, North Carolina State
University, Raleigh, NC 27695, USA \linebreak}

\begin{abstract}
We derive the generalization of Wigner's causality bounds and Bethe's integral
formula for the effective range parameter to arbitrary dimension and arbitrary
angular momentum. We also discuss the impact of these constraints on the
separation of low- and high-momentum scales and universality in low-energy
scattering. Some of our results were summarized earlier in a letter
publication. In this work, we present full derivations and several detailed examples.

\end{abstract}
\maketitle

\section{Introduction}

\begin{figure}[t]
\begin{center}
\includegraphics[
height=2.6299in,
width=2.6965in
]{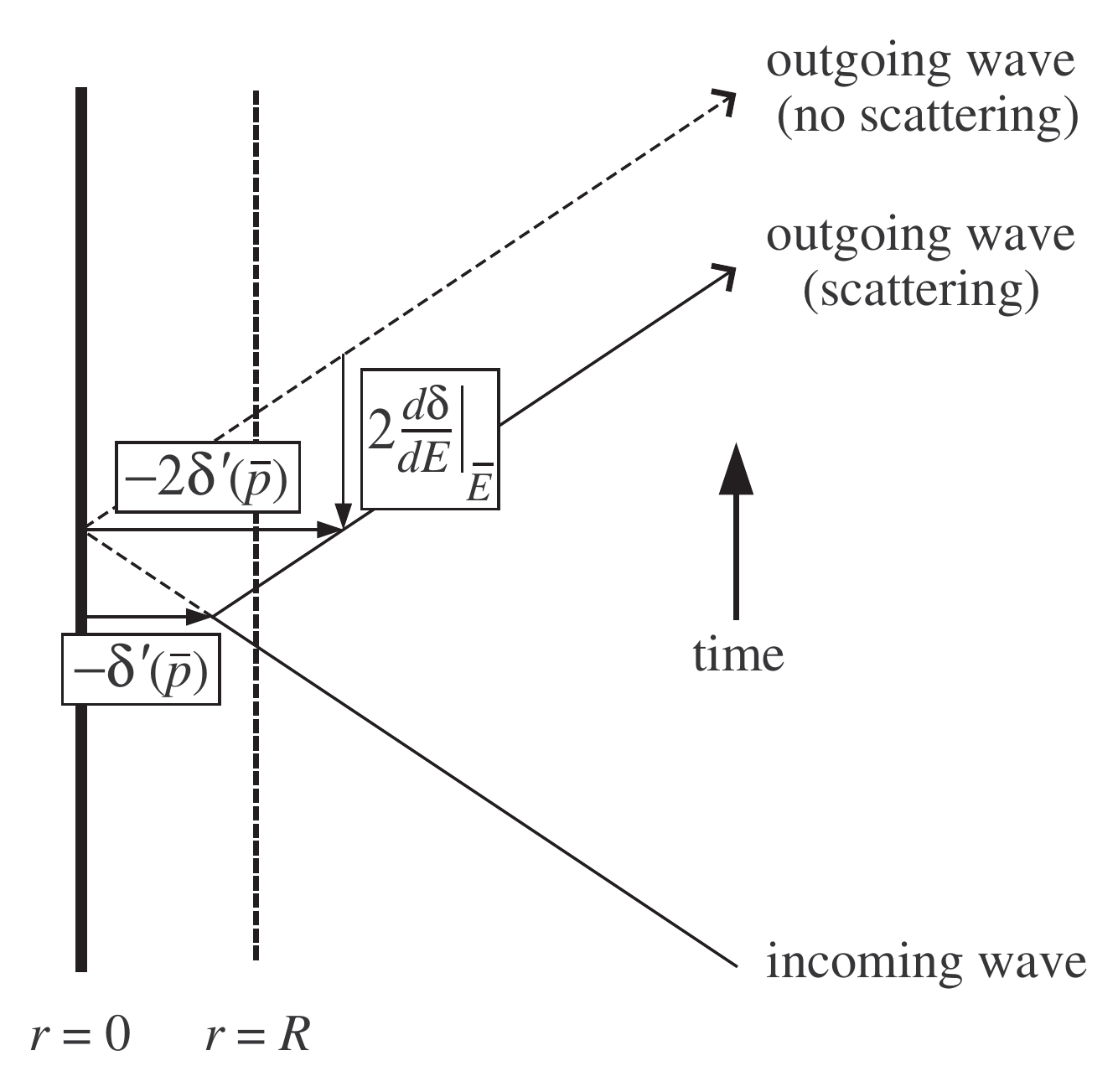}
\end{center}
\caption{Time versus radial distance. \ The scattered wavepacket is shifted in
distance by $-2\delta^{\prime}(\bar{p})$ and shifted in time by $2\left.
\frac{d\delta}{dE}\right\vert _{\bar{E}}$.}%
\label{time_delay}%
\end{figure}

Causality plays a fundamental role in physics. The principle that no action
can be observed before its cause puts important constraints on physical
theories. In classical electrodynamics, causality leads to the Kramers-Kronig
relations which relate the real and imaginary parts of the dielectric
constant. In quantum mechanics, causality requires that no scattered wave is
produced before the incident wave first reaches the scatterer. \ For
finite-range interactions the constraints of causality on elastic phase shifts
were first investigated by Wigner \cite{Wigner:1955a}. \ To illustrate the
underlying physics, we consider a wavepacket of outgoing spherical waves in
$d$ spatial dimensions,
\begin{equation}
f_{\text{out}}(r)=\int_{0}^{\infty}dp\;e^{ipr}\tilde{f}_{\text{out}}(p),
\end{equation}
where normalization factors and the $r^{-(d-1)/2}$ radial dependence are
absorbed into the definition of $f_{\text{out}}(r)$. \ When this wavepacket is
scattered, the $S$-matrix multiplies asymptotic outgoing states by a phase
factor $e^{2i\delta(p)}$, where $\delta(p)$ is the elastic phase shift. \ We
assume the momentum distribution $\tilde{f}_{\text{out}}(p)$ is sharply peaked
around some nonzero value $\bar{p}$. If $f_{\text{out}}^{\delta}(r)$ is the
scattered wavepacket, then%
\begin{align}
f_{\text{out}}^{\delta}(r)  &  =\int_{0}^{\infty}dp\;e^{ipr}e^{2i\delta
(p)}\tilde{f}_{\text{out}}(p)\nonumber\\
&  \approx e^{2i\delta(\bar{p})}e^{-2i\delta^{\prime}(\bar{p})\bar{p}%
}f_{\text{out}}\left[  r+2\delta^{\prime}(\bar{p})\right]  .
\end{align}
As a consequence, the wavepacket is shifted forward by $\Delta r=-2\delta
^{\prime}(\bar{p})$ relative to the wavepacket with no scattering. \ If we
consider the wavepacket as a function of time, the same shift can be
interpreted as a time shift or delay for the scattered wavepacket
\cite{Wigner:1955a},%
\begin{equation}
\Delta t=2\left.  \frac{d\delta}{dE}\right\vert _{\bar{E}},
\end{equation}
where $\bar{E}$ is the energy corresponding with $\bar{p}$. \ The radius shift
and time delay are sketched in Fig.~\ref{time_delay}.
%
A classical analysis of particle trajectories suggests that if the
interactions have a finite range $R$, then causality requires $-\delta
^{\prime}(\bar{p})\leq R$. \ While this argument is qualitatively correct, it
ignores the quantum mechanical spread of the wavepacket in space. \ The
precise causality bound and the consequences of the bound are the subject of
the present analysis.


In a previous letter, we have explored the impact of causality constraints on
low-energy universality \cite{Hammer:2009zh}. Low-energy universality can
appear when there is a large separation between the short-distance scale of
the interaction and the long-distance scales relevant to the physical system.
\ One example of low-energy universality is the unitarity limit. \ The strict
unitarity limit refers to an idealized system where the range of the
interaction is zero and the $S$-wave scattering length is infinite. In
experiment, the strict unitarity limit can not be reached because real systems
always have finite range interactions. Since finite range effects are
suppressed at low energies, one also refers to systems in the unitarity limit
if they have infinite scattering length but finite range. \ In nuclear
physics, cold dilute neutron matter is close to the unitarity limit since the
neutron-neutron scattering length is much larger than the range. \ Stable
many-body systems with infinite scattering length can be created in
experiments with two types of ultracold fermionic atoms using Feshbach resonances.

The implications of causality for universality in $S$-wave scattering are well
understood \cite{Phillips:1996ae}. In this case, the unitarity limit can
always be reached by tuning the scattering length to infinity. \ For reviews
of recent cold atom experiments exploring physics of the unitarity limit, see
Refs.~\cite{Koehler:2006A, Regal:2006thesis}. \ Theoretical overviews of
ultracold Fermi gases close to the unitarity limit and their numerical
simulations are given in \cite{Giorgini:2007a,Lee:2008fa}. \ A general review
of universality at large scattering length can be found in
\cite{Braaten:2004a}.

The implications of causality for large scattering length physics in higher
partial waves are more intricate. Because of causality, it is not always
possible to reach the unitarity limit by tuning external parameters
\cite{Barford:2002je,Hammer:2009zh}. Several experiments have investigated
strongly-interacting $P$-wave Feshbach resonances in $^{6}$Li and $^{40}$K
\cite{Regal:2003B, Ticknor:2004A, Zhang:2004A, Schunk:2005A, Gaebler:2007A}.
\ A key question is whether the physics of these strongly-interacting $P$-wave
systems is universal, and if so, what are the relevant low-energy parameters.
\ A positive answer to this question would provide a connection between the
atomic physics of $P$-wave Feshbach resonances and the nuclear physics of
$P$-wave alpha-neutron interactions in halo nuclei. \ Some progress in
addressing these questions has been made with low-energy models of $P$-wave
atomic interactions
\cite{Ohashi:2004A,Chevy:2004A,Gurarie:2005A,Levinsen:2007A,Gubbels:2007,Jonalasinio:2007A}
and $P$-wave alpha-neutron interactions
\cite{Bertulani:2002sz,Bedaque:2003wa,Nollett:2006su,Higa:2008rx}. \ A
renormalization group study showed that scattering should be weak in higher
partial waves unless there is a fine tuning of multiple parameters
\cite{Barford:2002je}. Complementary work was carried out by Ruiz Arriola and
collaborators. A discussion of the Wigner bound in the context of chiral
two-pion exchange can be found in \cite{PavonValderrama:2005wv} while
correlations between the scattering length and effective range related to the
Wigner bound were discussed in \cite{Cordon:2009pj}.

In Ref.~\cite{Hammer:2009zh}, we have addressed the question of
universality\ and provided expressions for the causality constraints in
arbitrary dimension $d$ and for arbitrary angular momentum $L$. \ Our analysis
applies to any finite-range interaction that is energy independent,
non-singular, and spin independent. \ Our results can be viewed as a
generalization of the analysis of Phillips and Cohen \cite{Phillips:1996ae},
who derived a Wigner bound for the $S$-wave effective range for short-range
interactions in three dimensions. In the current paper, we present full
derivations of Wigner's causality bounds \cite{Wigner:1955a} and Bethe's
integral formula \cite{Bethe:1949yr} for the effective range parameter to
arbitrary dimension $d$ and angular momentum $L$. \ The extension of Bethe's
integral formula for $d=3$ and $L>0$ was first derived by Madsen
\cite{Madsen:2002a}. \ We discuss the impact of these constraints on the
separation of low- and high-momentum scales and universality in low-energy
scattering using several detailed examples.

The paper is organized as follows. In Sec.~\ref{sec:prel}, we set the stage by
defining angular momentum, scattering phase shifts, and the effective range
expansion for the general case of $d$ spatial dimensions. The equation for the
radial wave function and the Wronskian identity for two solutions with
different energies are derived in Sec.~\ref{sec:radial}. This identity is used
in Sec.~\ref{sec:causa} to derive the causality bound on the effective range
for zero energy and the general bound for finite energies. In
Sec.~\ref{sec:causacons}, we discuss the impact of these bounds on low-energy
universality. In particular, we address the question of the unitarity limit.
The impact of the causality bounds is illustrated in detail in
Sec.~\ref{sec:impact} using three different examples: a spherical step
potential in $d$ dimensions, alpha-neutron scattering which corresponds to
exponentially bounded interactions, and the long-range van der Waals
interaction. We end with a summary and outlook in Sec.~\ref{sec:sum}. Finally,
in the Appendices we explicitly check the Wronskian identities for all
possible values of $d$ and $L$ and demonstrate the equivalence of our
causality bound with Wigner's original bound on the energy derivative of the
phase shift.


\section{Preliminaries}

\label{sec:prel}

Our goal is to generalize Bethe's integral formula for the effective range
parameter and the related causality bound for arbitrary $d$ and $L$. We start
with some general definitions.

We consider two non-relativistic spinless particles in $d$ dimensions with a
rotationally-invariant two-body interaction. The interaction is assumed to be
energy independent and have a finite range $R$ beyond which the particles are
non-interacting.
$\ $Let $\mu$ be the reduced mass and $p^{2}/(2\mu)$ be the total energy.
\ For $d>1$ angular momentum is specified by $d-1$ integer quantum numbers
\cite{Avery:1982,Bolle:1984A}
\begin{equation}
\mathbf{L=}\left\{  M_{1},\cdots,M_{d-1}\right\}  ,
\end{equation}
satisfying%
\begin{equation}
\left\vert M_{1}\right\vert \leq M_{2}\leq\cdots\leq M_{d-2}\leq M_{d-1}.
\end{equation}
We let $L$ label the absolute value of the top-level quantum number,
$\left\vert M_{d-1}\right\vert $. \ For example when $d=3$, $M_{1}$ is the
angular momentum projection $M=-L,\ldots, L$ and $M_{2}=L=0,1,2,\ldots$ is the
total angular momentum. In $d=2$, there is only one angular momentum quantum
number $|M_{1}|=L=0,1,2,\ldots$. The case of one spatial dimension is special
since continuous rotations do not exist.
\ Instead of rotational invariance, the key symmetry in one dimension is
invariance under parity. \ We assume a parity-symmetric interaction and write
$\mathbf{L=}L=0$ for even parity and $\mathbf{L=}L=1$ for odd parity. \ In the
following all results we derive for rotationally-invariant interactions in
$d>1$ are also valid for parity-symmetric interactions in $d=1$.

We analyze the two-body system in the center-of-mass frame using units with
$\hbar=1$ for convenience. The full wave function for reduced mass $\mu$ and
energy $p^{2}/(2\mu),$ can be separated into a radial part and an angular part
via
\begin{equation}
\Psi^{(p)}_{L,d}(\mathbf{r})=R_{L,d}^{(p)}(r) Y_{\mathbf{L}}(\hat{r}),
\end{equation}
where the $Y_{\mathbf{L}}(\hat{r})$ are hyperspherical harmonics. The
hyperspherical harmonics in $d$ dimensions satisfy the orthogonality
condition,%
\begin{equation}
\int Y_{\mathbf{L}}^{\ast}(\hat{r})Y_{\mathbf{L}^{\prime}}(\hat{r}%
)\ d\Omega_{d}=\delta_{\mathbf{L}^{\prime},\mathbf{L}},
\end{equation}
where $\Omega_{d}$ is the solid angle and the sum rule \cite{Avery:1982},%
\begin{equation}
\sum_{\mathbf{L}\text{, }L\text{ fixed}}Y_{\mathbf{L}}^{\ast}(\hat
{r})Y_{\mathbf{L}}(\hat{r}^{\prime})=\frac{\left(  d+2L-2\right)  \left(
d-4\right)  !!}{N_{d}}C_{L}^{d/2-1}(\hat{r}\cdot\hat{r}^{\prime}).
\label{projection_id}%
\end{equation}
Here $C_{L}^{d/2-1}$ is a Gegenbauer polynomial, and the normalization factor
$N_{d}$ is given by,%
\begin{equation}
N_{d}=\frac{(d-2)!!2\pi^{d/2}}{\Gamma\left(  \frac{d}{2}\right)  }.
\end{equation}
For $d=2$ the expressions are defined in the limit $d\rightarrow2$.

We can check that these give the expected sum rules for $d\leq3$. \ In one
dimension we get even and odd parity functions of $\hat{r}\cdot\hat{r}%
^{\prime}$,%
\begin{equation}
\left(  d+2L-2\right)  \left(  d-4\right)  !!C_{L}^{d/2-1}(\hat{r}\cdot\hat
{r}^{\prime})=\left\{
\begin{array}
[c]{c}%
1\text{ for }L=0\\
\hat{r}\cdot\hat{r}^{\prime}\text{ for }L=1.
\end{array}
\right.
\end{equation}
For two dimensions, we have a sum of modes $e^{\pm iL\theta}$ for $\theta
=\cos^{-1}\left(  \hat{r}\cdot\hat{r}^{\prime}\right)  $,%
\begin{equation}
\lim_{d\rightarrow2}\left(  d+2L-2\right)  \left(  d-4\right)  !!C_{L}%
^{d/2-1}(\hat{r}\cdot\hat{r}^{\prime})=\left\{
\begin{array}
[c]{c}%
1\text{ for }L=0\\
2\cos\left[  L\cos^{-1}\left(  \hat{r}\cdot\hat{r}^{\prime}\right)  \right]
\text{ for }L>0.
\end{array}
\right.
\end{equation}
\ In three dimensions, we recover the Legendre polynomials$,$%
\begin{equation}
\left(  d+2L-2\right)  \left(  d-4\right)  !!C_{L}^{d/2-1}(\hat{r}\cdot\hat
{r}^{\prime})=(2L+1)P_{L}(\hat{r}\cdot\hat{r}^{\prime}).
\end{equation}


The scattering phase shifts are directly related to the elastic scattering
amplitude $f_{L,d}(p)$, where%
\begin{equation}
f_{L,d}(p)\propto\frac{p^{2L}}{p^{2L+d-2}\cot\delta_{L,d}(p)-ip^{2L+d-2}}.
\label{f_L}%
\end{equation}
In addition to having finite range, we assume also that $p^{2L+d-2}\cot
\delta_{L,d}(p)$ does not diverge at $p=0$ and the interaction is not too
singular at short distances. \ Specifically, we require that the effective
range expansion defined below in Eq.~(\ref{ere}) converges for sufficiently
small $p$. Moreover, the reduced radial wave function $u_{L,d}^{(p)}$ defined
in Eq.~(\ref{eq:red_wf}) must satisfy that $\frac{d}{dr}u_{L,d}^{(p)}$ is
finite and $u_{L,d}^{(p)}$ vanishes as $r\rightarrow0$. \ In
\cite{Bolle:1984A}, these short-distance regularity conditions are shown to be
fulfilled for interactions arising from a static potential$,$%
\begin{equation}
W(r,r^{\prime})=V(r)\delta(r-r^{\prime}),
\end{equation}
provided that $V(r)=O(r^{-2+\epsilon})$ as $r\rightarrow0$ for positive
$\epsilon$. \ In our discussion, however, we make no assumption that the
interactions correspond to a local potential.

The effective range expansion for general dimension $d$ and angular momentum
$L$ is%
\begin{align}
&  p^{2L+d-2}\left[  \cot\delta_{L,d}(p)-\delta_{(d\operatorname{mod}%
2),0}\frac{2}{\pi}\ln\left(  p\rho_{L,d}\right)  \right] \nonumber\\
&  =-\frac{1}{a_{L,d}}+\frac{1}{2}r_{L,d}p^{2}+\sum_{n=0}^{\infty}%
(-1)^{n+1}\mathcal{P}_{L,d}^{(n)}p^{2n+4}. \label{ere}%
\end{align}
The term $\delta_{(d\operatorname{mod}2),0}$ is $0$ for odd $d$ and $1$ for
even $d$. \ $a_{L,d}$ is the scattering parameter,\footnote{For $S$-wave
scattering in three spatial dimensions, $a_{0,3}$ has dimensions of length and
is usually called \textit{scattering length}.} $r_{L,d}$ is the effective
range parameter, and $\mathcal{P}_{L,d}^{(n)}$ are the $n^{\text{th}}$-order
shape parameters. $\ \rho_{L,d}$ is an arbitrary length scale that can be
scaled to any nonzero value. \ The rescaling results in a shift of the
dimensionless coefficient of $p^{2L+d-2}$ on the right-hand side of
Eq.~(\ref{ere}), and we define $\bar{\rho}_{L,d}$ as the special value for
$\rho_{L,d}$ where this coefficient is zero.

Throughout our discussion we use the examples of $S$-wave and $P$-wave
scattering in three dimensions to illustrate general formulas. \ For $d=3$ and
$L=0$ the scattering amplitude is%
\begin{equation}
f_{0,3}(p)\propto\frac{1}{p\cot\delta_{0,3}(p)-ip},
\end{equation}
and the effective range expansion is%
\begin{equation}
p\cot\delta_{0,3}(p)=-\frac{1}{a_{0,3}}+\frac{1}{2}r_{0,3}p^{2}+\sum
_{n=0}^{\infty}(-1)^{n+1}\mathcal{P}_{0,3}^{(n)}p^{2n+4}.
\end{equation}
For $d=3$ and $L=1$ we have%
\begin{equation}
f_{1,3}(p)\propto\frac{p^{2}}{p^{3}\cot\delta_{1,3}(p)-ip^{3}}%
\end{equation}
and%
\begin{equation}
p^{3}\cot\delta_{1,3}(p)=-\frac{1}{a_{1,3}}+\frac{1}{2}r_{1,3}p^{2}+\sum
_{n=0}^{\infty}(-1)^{n+1}\mathcal{P}_{1,3}^{(n)}p^{2n+4}.
\end{equation}


\section{Radial equation and Wronskian identity}

\label{sec:radial}

The next step is the derivation of the Wronskian identity for the solutions of
the radial Schr\"{o}dinger equation. The causality bound then follows directly
from this identity.

The interaction is assumed to have finite range $R$ beyond which the particles
are non-interacting. \ With the interaction written as a real symmetric
operator with kernel $W(r,r^{\prime})$, the radial Schr\"{o}dinger equation is%
\begin{align}
p^{2}R_{L,d}^{(p)}(r)  &  =-\frac{1}{r^{d-1}}\frac{d}{dr}\left[  r^{d-1}%
\frac{d}{dr}R_{L,d}^{(p)}(r)\right]  +\frac{L(L+d-2)}{r^{2}}R_{L,d}%
^{(p)}(r)\nonumber\\
&  +2\mu\int_{0}^{R}dr^{\prime}W(r,r^{\prime})u_{L,d}^{(p)}(r^{\prime}).
\end{align}
We rescale the radial wave function $R_{L,d}^{(p)}(r)$ as
\begin{equation}
u_{L,d}^{(p)}(r)=\left(  pr\right)  ^{(d-1)/2}R_{L,d}^{(p)}(r),
\label{eq:red_wf}%
\end{equation}
and obtain
\begin{align}
p^{2}u_{L,d}^{(p)}(r)  &  =-\frac{d^{2}}{dr^{2}}u_{L,d}^{(p)}(r)+\frac{\left(
2L+d-1\right)  \left(  2L+d-3\right)  }{4r^{2}}u_{L,d}^{(p)}(r)\nonumber\\
&  +2\mu\int_{0}^{R}dr^{\prime}W(r,r^{\prime})u_{L,d}^{(p)}(r^{\prime}).
\end{align}

The normalization of $u_{L,d}^{(p)}(r)$ is chosen so that for $r\geq R$,%
\begin{align}
u_{L,d}^{(p)}(r)  &  =\sqrt{\frac{pr\pi}{2}}p^{L+d/2-3/2}\left[  \cot
\delta_{L,d}(p)J_{L+d/2-1}(pr)-Y_{L+d/2-1}(pr)\right] \nonumber\\
&  =p^{L+d/2-3/2}\left[  \cot\delta_{L,d}(p)\times S_{L+d/2-3/2}%
(pr)+C_{L+d/2-3/2}(pr)\right]  . \label{u_Ld}%
\end{align}
Here $J_{\alpha}$ and $Y_{\alpha}$ are the Bessel functions of the first and
second kind, $S_{\alpha}$ and $C_{\alpha}$ are the Riccati-Bessel functions of
the first and second kind, and $\delta_{L,d}(p)$ is the phase shift for
partial wave $L$. Our conventions for the Bessel functions and Riccati-Bessel
functions are given in Appendix \ref{app:b_Ld}. \ In the following, we use the
notation $u_{L,d}^{(0)}(r)$ as shorthand for the limit $p\rightarrow0$,%
\begin{equation}
u_{L,d}^{(0)}(r)=\lim_{p\rightarrow0}u_{L,d}^{(p)}(r)\text{.}%
\end{equation}

For our first example, $d=3$ and $L=0$, the rescaled radial wave function for
$r\geq R$ is%
\begin{align}
u_{0,3}^{(p)}(r)  &  =\sqrt{\frac{pr\pi}{2}}\left[  \cot\delta_{0,3}%
(p)J_{1/2}(pr)-Y_{1/2}(pr)\right] \nonumber\\
&  =\frac{\sin\left[  pr+\delta_{0,3}(p)\right]  }{\sin\left[  \delta
_{0,3}(p)\right]  }.
\end{align}
This satisfies the radial equation,%
\begin{equation}
p^{2}u_{0,3}^{(p)}(r)=-\frac{d^{2}}{dr^{2}}u_{0,3}^{(p)}(r).
\end{equation}
For $d=3$ and $L=1$, we find%
\begin{align}
u_{1,3}^{(p)}(r)  &  =p\sqrt{\frac{pr\pi}{2}}\left[  \cot\delta_{1,3}%
(p)J_{3/2}(pr)-Y_{3/2}(pr)\right] \nonumber\\
&  =\frac{\sin\left[  pr+\delta_{1,3}(p)\right]  -pr\cos\left[  pr+\delta
_{1,3}(p)\right]  }{r\sin\left[  \delta_{1,3}(p)\right]  }.
\end{align}
In this case%
\begin{equation}
p^{2}u_{1,3}^{(p)}(r)=\left(  -\frac{d^{2}}{dr^{2}}+\frac{2}{r^{2}}\right)
u_{1,3}^{(p)}(r).
\end{equation}


We choose two values for the momenta $p_{A}$ and $p_{B}$. \ We use the
shorthand $\delta_{A}=\delta_{L,d}(p_{A})$, $\delta_{B}=\delta_{L,d}(p_{B})$.
\ Similarly $u_{A}(r)=u_{L,d}^{(p_{A})}(r)$ and $u_{B}(r)=u_{L,d}^{(p_{B}%
)}(r)$. \ Therefore%
\begin{equation}
p_{A}^{2}u_{A}(r)=-u_{A}^{\prime\prime}(r)+\frac{\left(  2L+d-1\right)
\left(  2L+d-3\right)  }{4r^{2}}u_{A}(r)+2\mu\int_{0}^{R}dr^{\prime
}W(r,r^{\prime})u_{A}(r^{\prime}),\label{uA}%
\end{equation}%
\begin{equation}
p_{B}^{2}u_{B}(r)=-u_{B}^{\prime\prime}(r)+\frac{\left(  2L+d-1\right)
\left(  2L+d-3\right)  }{4r^{2}}u_{B}(r)+2\mu\int_{0}^{R}dr^{\prime
}W(r,r^{\prime})u_{B}(r^{\prime}).\label{uB}%
\end{equation}
We now multiply Eq.~(\ref{uA}) by $u_{B}$, multiply Eq.~(\ref{uB}) by $u_{A}$,
and subtract the two,%
\begin{align}
&  \left(  p_{A}^{2}-p_{B}^{2}\right)  u_{A}(r)u_{B}(r)\nonumber\\
&  =-u_{B}(r)u_{A}^{\prime\prime}(r)+u_{A}(r)u_{B}^{\prime\prime
}(r)\nonumber\\
&  +2\mu\int_{0}^{R}dr^{\prime}\left[  u_{B}(r)W(r,r^{\prime})u_{A}(r^{\prime
})-u_{A}(r)W(r,r^{\prime})u_{B}(r^{\prime})\right]  .
\end{align}
Integrating from radius $\rho$ to some radius $r\geq R$, we get%
\begin{align}
&  \left(  p_{A}^{2}-p_{B}^{2}\right)  \int_{\rho}^{r}dr^{\prime}%
\,u_{A}(r^{\prime})u_{B}(r^{\prime})\nonumber\\
&  =-\int_{\rho}^{r}dr^{\prime}\,u_{B}(r^{\prime})u_{A}^{\prime\prime
}(r^{\prime})+\int_{\rho}^{r}dr^{\prime}\,u_{A}(r^{\prime})u_{B}^{\prime
\prime}(r^{\prime})\nonumber\\
&  +2\mu\int_{\rho}^{R}dr\int_{0}^{R}dr^{\prime}\left[  u_{B}(r)W(r,r^{\prime
})u_{A}(r^{\prime})-u_{A}(r)W(r,r^{\prime})u_{B}(r^{\prime})\right]  ,
\end{align}
and therefore%
\begin{align}
&  \left(  p_{B}^{2}-p_{A}^{2}\right)  \int_{\rho}^{r}dr^{\prime}%
\,u_{A}(r^{\prime})u_{B}(r^{\prime})\nonumber\\
&  =\left.  \left(  u_{B}u_{A}^{\prime}-u_{A}u_{B}^{\prime}\right)
\right\vert _{\rho}^{r}\nonumber\\
&  -2\mu\int_{\rho}^{R}dr\int_{0}^{R}dr^{\prime}\left[  u_{B}(r)W(r,r^{\prime
})u_{A}(r^{\prime})-u_{A}(r)W(r,r^{\prime})u_{B}(r^{\prime})\right]  .
\end{align}

By assumption the interaction is sufficiently well-behaved at the origin and
admits a regular solution. \ In particular,%
\begin{equation}
\lim_{\rho\rightarrow0^{+}}u_{B}\left(  \rho\right)  u_{A}^{\prime}\left(
\rho\right)  =\lim_{\rho\rightarrow0^{+}}u_{A}\left(  \rho\right)
u_{B}^{\prime}\left(  \rho\right)  =0.
\end{equation}
So we have
\begin{equation}
u_{B}\left(  r\right)  u_{A}^{\prime}\left(  r\right)  -u_{A}\left(  r\right)
u_{B}^{\prime}\left(  r\right)  =\left(  p_{B}^{2}-p_{A}^{2}\right)  \int
_{0}^{r}dr^{\prime}\,u_{A}(r^{\prime})u_{B}(r^{\prime}).
\label{Wronskian_integral1}%
\end{equation}
The left-hand side of Eq.~(\ref{Wronskian_integral1}) corresponds with the
Wronskian of $u_{B}$ and $u_{A}$, $W[u_{B},u_{A}](r)$, and so we have%
\begin{equation}
W[u_{B},u_{A}](r)=\left(  p_{B}^{2}-p_{A}^{2}\right)  \int_{0}^{r}dr^{\prime
}\,u_{A}(r^{\prime})u_{B}(r^{\prime}). \label{Wronskian_integral2}%
\end{equation}

In the following, it will be useful to rearrange Eq.~(\ref{u_Ld}) and express
$u_{L,d}^{(p)}(r)$ for $r\geq R$ in terms of functions $s(p,r)$ and $c(p,r)$
so that%
\begin{equation}
u_{L,d}^{(p)}(r)=p^{2L+d-2}\left[  \cot\delta_{L,d}(p)-\delta
_{(d\operatorname{mod}2),0}\frac{2}{\pi}\ln\left(  p\rho_{L,d}\right)
\right]  s(p,r)+c(p,r). \label{standard_form}%
\end{equation}
\ Later in our discussion we derive $s(p,r)$ and $c(p,r)$ for each of the
possible cases for $2L+d$, and show that both functions are analytic in
$p^{2}$. \ The first two series coefficients will be useful,%
\begin{equation}
s(p,r)=s_{0}(r)+s_{2}(r)p^{2}+O(p^{4}),
\end{equation}%
\begin{equation}
c(p,r)=c_{0}(r)+c_{2}(r)p^{2}+O(p^{4}).
\end{equation}
Combining with the effective range expansion, we find that
\begin{equation}
u_{L,d}^{(p)}(r)=-\frac{1}{a_{L,d}}s_{0}(r)+c_{0}(r)+\left[  \frac{1}%
{2}r_{L,d}s_{0}(r)-\frac{1}{a_{L,d}}s_{2}(r)+c_{2}(r)\right]  p^{2}+O(p^{4}).
\end{equation}
The Wronskian $W[u_{B},u_{A}](r)$ for $r\geq R$ is then%
\begin{align}
W[u_{B},u_{A}](r)  &  =\left(  p_{B}^{2}-p_{A}^{2}\right)  \left\{  \frac
{1}{2}r_{L,d}W[s_{0},c_{0}](r)+\left(  \frac{1}{a_{L,d}}\right)  ^{2}%
W[s_{2},s_{0}](r)\right. \nonumber\\
&  \left.  -\frac{1}{a_{L,d}}W[s_{2},c_{0}](r)-\frac{1}{a_{L,d}}W[c_{2}%
,s_{0}](r)+W[c_{2},c_{0}](r)\right\} \nonumber\\
&  +O(p_{A}^{4})+O(p_{B}^{4})\text{.} \label{Wronskian_expansion}%
\end{align}
Starting from the Wronskian integral formula Eq.~(\ref{Wronskian_integral2}),
we set $p_{A}=0$ and take the limit $p=p_{B}\rightarrow0$. \ When combined
with the expansion in Eq.~(\ref{Wronskian_expansion}) we find that for $r\geq
R$,%
\begin{equation}
-r_{L,d}W[s_{0},c_{0}](r)=b_{L,d}(r)-2\int_{0}^{r}dr^{\prime}\,\left[
u_{L,d}^{(0)}(r^{\prime})\right]  ^{2}, \label{r_Ld_general}%
\end{equation}
where%
\begin{equation}
b_{L,d}(r)=2W[c_{2},c_{0}](r)-\frac{2}{a_{L,d}}\left\{  W[s_{2},c_{0}%
](r)+W[c_{2},s_{0}](r)\right\}  +\frac{2}{a_{L,d}^{2}}W[s_{2},s_{0}](r).
\label{b_Ld_general}%
\end{equation}
A fundamental result for second-order differential equations known as Abel's
differential equation identity \cite{Abel:1829} implies that the Wronskian of
$s(p,r)$ and $c(p,r)$ is independent of $r$. \ Given our choice of
normalization, the Wronskian of $s(p,r)$ and $c(p,r)$ is $-1$ for all $p$.
\ \ This implies that%
\begin{equation}
W[s_{0},c_{0}](r)=-1, \label{eq:idW1}%
\end{equation}%
\begin{equation}
W[s_{2},c_{0}](r)=W[c_{2},s_{0}](r). \label{eq:idW2}%
\end{equation}
In Appendix \ref{app:b_Ld}, we check explicitly that these identities hold in
all cases and derive explicit expressions for $b_{L,d}(r)$.

For our first example, $d=3$ and $L=0$, the functions $s(p,r)$ and $c(p,r)$
are%
\begin{equation}
s(p,r)=\frac{\sin pr}{p},
\end{equation}%
\begin{equation}
c(p,r)=\cos pr.
\end{equation}
The low-momentum expansions of these functions are%
\begin{equation}
s(p,r)=r-\frac{r^{3}}{6}p^{2}+O(p^{4}),
\end{equation}%
\begin{equation}
c(p,r)=1-\frac{r^{2}}{2}p^{2}+O(p^{4}).
\end{equation}
From Eq.~(\ref{b_Ld_general}) the Wronksians of the expansion coefficients
give%
\begin{equation}
b_{0,3}(r)=2r-\frac{2r^{2}}{a_{0,3}}+\frac{2r^{3}}{3a_{0,3}^{2}}.
\end{equation}

For our second example, $d=3$ and $L=1$, the functions $s(p,r)$ and $c(p,r)$
are%
\begin{equation}
s(p,r)=\frac{1}{p^{2}}\left(  \frac{\sin pr}{pr}-\cos pr\right)  ,
\end{equation}%
\begin{equation}
c(p,r)=p\left(  \frac{\cos pr}{pr}+\sin pr\right)  .
\end{equation}
In this case the low-momentum expansions are%
\begin{equation}
s(p,r)=\frac{r^{2}}{3}-\frac{r^{4}}{30}p^{2}+O(p^{4}),
\end{equation}%
\begin{equation}
c(p,r)=\frac{1}{r}+\frac{r}{2}p^{2}+O(p^{4}).
\end{equation}
The Wronksians of the coefficients lead to the result%
\begin{equation}
b_{1,3}(r)=-\frac{2}{r}-\frac{2r^{2}}{3a_{1,3}}+\frac{2r^{5}}{45a_{1,3}^{2}}.
\end{equation}

\section{Causality bounds}

\label{sec:causa}

We are now in the position to write down the causality bound for the effective
range $r_{L,d}$. Using the Wronskian identities, Eqs.~(\ref{eq:idW1},
\ref{eq:idW2}), we can simplify Eqs.~(\ref{r_Ld_general}, \ref{b_Ld_general})
to
\begin{equation}
r_{L,d}=b_{L,d}(r)-2\int_{0}^{r}dr^{\prime}\,\left[  u_{L,d}^{(0)}(r^{\prime
})\right]  ^{2}, \label{r_Ld}%
\end{equation}%
\begin{equation}
b_{L,d}(r)=2W[c_{2},c_{0}](r)-\frac{4}{a_{L,d}}W[s_{2},c_{0}](r)+\frac
{2}{a_{L,d}^{2}}W[s_{2},s_{0}](r). \label{b_Ld_simple}%
\end{equation}
These equations hold for any $r\geq R$. In Appendix \ref{app:b_Ld}, we derive
explicit expressions for the quantity $b_{L,d}(r)$ for all relevant
combinations of $d$ and $L$. In particular, we find for $2L+d=2$:%
\begin{equation}
b_{L,d}(r)=\frac{2r^{2}}{\pi}\left\{  \left[  \ln\left(  \frac{r}{2\rho_{L,d}%
}\right)  +\gamma-\frac{1}{2}+\frac{\pi}{2a_{L,d}}\right]  ^{2}+\frac{1}%
{4}\right\}  , \label{eq:b_2Ld2}%
\end{equation}
for $2L+d=4$:%
\begin{equation}
b_{L,d}(r)=\frac{4}{\pi}\left[  \ln\left(  \frac{r}{2\rho_{L,d}}\right)
+\gamma\right]  -\frac{4}{a_{L,d}}\left(  \frac{r}{2}\right)  ^{2}+\frac{\pi
}{a_{L,d}^{2}}\left(  \frac{r}{2}\right)  ^{4}\text{,} \label{eq:b_2Ld4}%
\end{equation}
and when $2L+d$ is any positive odd integer or any even integer $\geq6$:%
\begin{align}
b_{L,d}(r)  &  =-\frac{2\Gamma(L+\frac{d}{2}-2)\Gamma(L+\frac{d}{2}-1)}{\pi
}\left(  \frac{r}{2}\right)  ^{-2L-d+4}\nonumber\\
&  -\frac{4}{L+\frac{d}{2}-1}\frac{1}{a_{L,d}}\left(  \frac{r}{2}\right)
^{2}\nonumber\\
&  +\frac{2\pi}{\Gamma(L+\frac{d}{2})\Gamma(L+\frac{d}{2}+1)}\frac{1}%
{a_{L,d}^{2}}\left(  \frac{r}{2}\right)  ^{2L+d}. \label{2Ld_odd}%
\end{align}
Since the integrand on the right-hand side of Eq.~(\ref{r_Ld}) is positive
semi-definite, $r_{L,d}$ satisfies the upper bound%
\begin{equation}
r_{L,d}\leq b_{L,d}(r) \label{eq:causa}%
\end{equation}
for any $r\geq R$.  Eq.~(\ref{eq:causa}) together with
Eqs.~(\ref{eq:b_2Ld2}, \ref{eq:b_2Ld4}, \ref{2Ld_odd}) constitutes the
generalization of the causality bound on the effective range for arbitrary
dimension $d$ and angular momentum $L$ \cite{Hammer:2009zh}.

We can extend our causality bound to nonzero values of $p$. \ Consider any
$p\neq0$. \ If $\sin\delta_{L,d}(p)\neq0$, then $W[u_{B},u_{A}](r)$ for $r\geq
R$ is an analytic function of $p_{A}$ and $p_{B}$ in a neighborhood of $p$.
\ In this neighborhood, we can also consider $W[u_{B},u_{A}](r)$ as an
analytic function of the variables $p_{B}^{2}-p_{A}^{2}$ and $p_{B}^{2}%
+p_{A}^{2}$. \ Since $W[u_{B},u_{A}](r)$ is antisymmetric with respect to
$p_{A}$ and $p_{B}$, it is an odd function of $p_{B}^{2}-p_{A}^{2}$. \ Hence%
\begin{equation}
\frac{W[u_{B},u_{A}](r)}{p_{B}^{2}-p_{A}^{2}}%
\end{equation}
is also analytic with respect to $p_{B}^{2}-p_{A}^{2}$ and $p_{B}^{2}%
+p_{A}^{2}$. \ Taking the limit $p_{B}\rightarrow p_{A}$, we find that
\begin{equation}
\lim_{p_{B}\rightarrow p_{A}}\frac{W[u_{B},u_{A}](r)}{p_{B}^{2}-p_{A}^{2}%
}=\int_{0}^{r}dr^{\prime}\,u_{A}^{2}(r^{\prime}).
\label{nonthreshold_integral}%
\end{equation}
Since the right-hand side is non-negative, we conclude that for any $p_{A}%
\neq0$ such that $\sin\delta_{L,d}(p_{A})\neq0$,%
\begin{equation}
\lim_{p_{B}\rightarrow p_{A}}\frac{W[u_{B},u_{A}](r)}{p_{B}^{2}-p_{A}^{2}}%
\geq0. \label{nonthreshold_causality_bound}%
\end{equation}
This is the generalization of the causality bound for nonzero $p$. The
equivalence of our causality bound with Wigner's original bound on the energy
derivative of the phase shift~\cite{Wigner:1955a} is demonstrated in Appendix
\ref{app:equiv}.

\section{Causality constraints on low-energy universality}

\label{sec:causacons}

We now discuss the impact of the causality constraints from
Eqs.~(\ref{eq:causa}, \ref{eq:b_2Ld2}, \ref{eq:b_2Ld4}, \ref{2Ld_odd}) on
low-energy universality.

\begin{figure}[ptb]
\begin{center}
\includegraphics[
height=3.3918in,
width=4.3275in
]{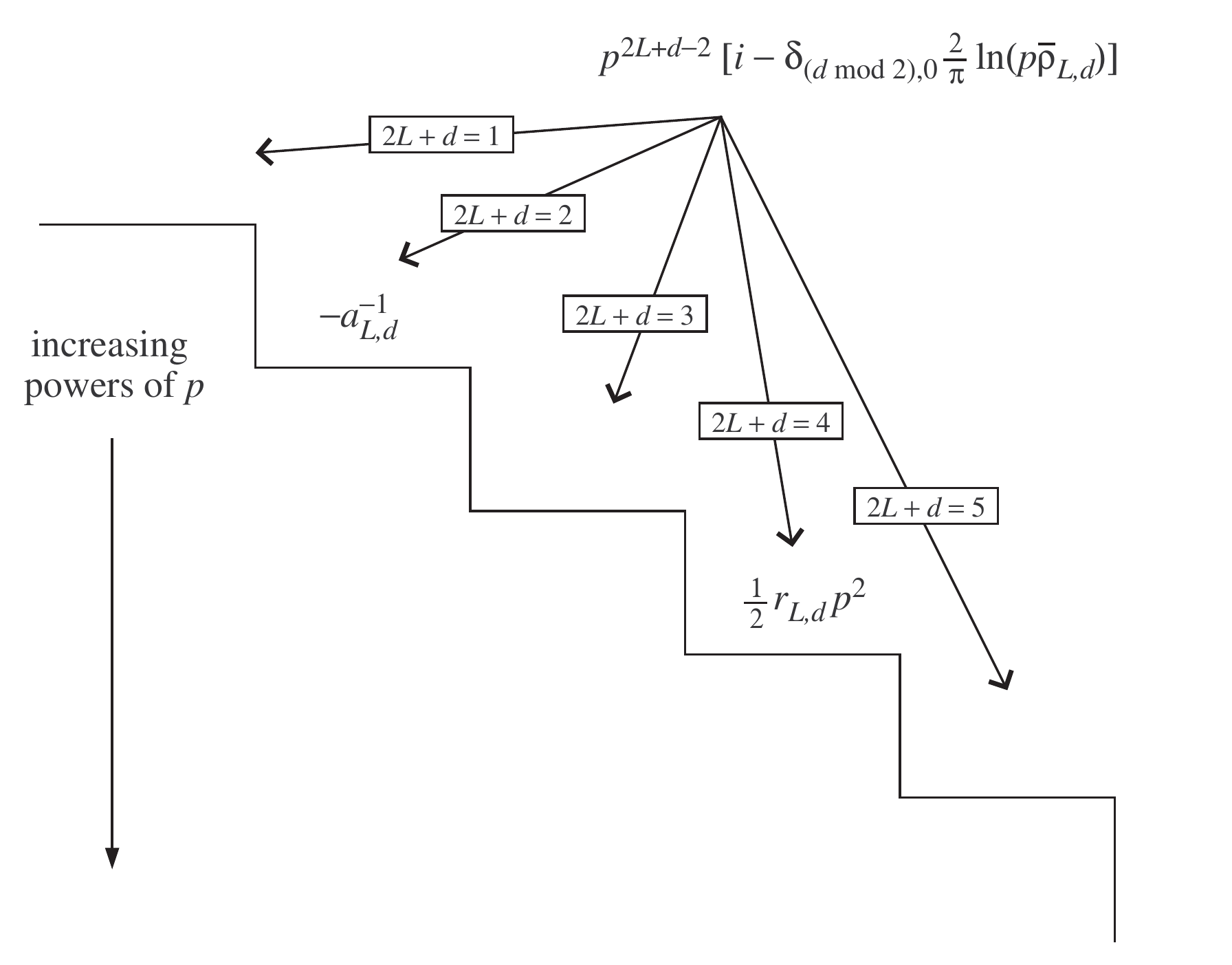}
\end{center}
\caption{Hierarchy of terms in the effective range expansion as $p\rightarrow
0$.}%
\label{staircase}%
\end{figure}

We consider the scattering amplitude in the low-energy limit $p\rightarrow0$
while keeping the interaction range $R$ fixed. \ Let $\alpha_{L,d}(p)$
describe an effective scattering parameter,%
\begin{equation}
\alpha_{L,d}(p)=\frac{-1}{p^{2L+d-2}\cot\delta_{L,d}(p)-ip^{2L+d-2}}.
\end{equation}
The scattering amplitude is proportional to $\alpha_{L,d}(p)$ times a factor
of $p^{2L}$ from the angular momentum projection,%
\begin{equation}
f_{L,d}(p)\propto p^{2L}\alpha_{L,d}(p)\text{.}%
\end{equation}
In the limit $p\rightarrow0$, the hierarchy of terms in the effective range
expansion depends on the value of $2L+d$. \ This hierarchy is sketched in
Fig.~\ref{staircase}. In particular, the effective range parameter is as
important at low energies as the unitarity contribution for $2L+d=4$ and
becomes more important for $2L+d \geq5$. This implies that the scale-invariant
unitarity limit can not be reached in those cases because the Wigner bound
prevents the effective range from being tuned to zero. In the following, we
discuss the various cases in detail.

For $2L+d=1$, we find that%
\begin{equation}
\alpha_{L,d}(p)=-ip+a_{L,d}^{-1}p^{2}+O(p^{4}).
\end{equation}
As $p\rightarrow0$, the effective scattering parameter has a scale-invariant
weak-coupling limit%
\begin{equation}
\alpha_{L,d}(p)\approx-ip\rightarrow0,
\end{equation}
with the dimensionful parameter, $a_{L,d}^{-1}$, determining the leading
correction to scale-invariant physics.

For $2L+d=2$, we find%
\begin{equation}
\alpha_{L,d}(p)=-\frac{\pi}{2\ln\left(  -ip\bar{\rho}_{L,d}\right)  }%
+O(p^{2}/\ln^{2}p),
\end{equation}
where $\bar{\rho}_{L,d}$ denotes the special value for $\rho_{L,d}$ that makes
the inverse scattering parameter $1/a_{L,d}$ on the right-hand side of
Eq.~(\ref{ere}) equal to zero. As $p\rightarrow0$, the effective scattering
parameter has a logarithmic weak-coupling limit%
\begin{equation}
\alpha_{L,d}(p)\approx-\frac{\pi}{2\ln\left(  -ip\bar{\rho}_{L,d}\right)
}\rightarrow0,
\end{equation}
parameterized by the length parameter $\bar{\rho}_{L,d}$.

For $2L+d\geq3,$ it is more convenient to consider the inverse effective
scattering parameter. \ For $2L+d=3,$%
\begin{equation}
\alpha_{L,d}^{-1}(p)=a_{L,d}^{-1}+ip+O(p^{2}).
\end{equation}
If we fine-tune the interaction so that $a_{L,d}^{-1}=0$, then as
$p\rightarrow0$ the effective scattering parameter has a scale-invariant
strong-coupling limit%
\begin{equation}
\alpha_{L,d}^{-1}(p)\approx ip\rightarrow0.
\end{equation}
For $d=3$ and $L=0$, this is the physics of the unitarity limit in three dimensions.

For $2L+d=4$, we have%
\begin{equation}
\alpha_{L,d}^{-1}(p)=a_{L,d}^{-1}-\frac{2}{\pi}p^{2}\ln\left(  -ip\bar{\rho
}_{L,d}\right)  +O(p^{4}),
\end{equation}
where $\bar{\rho}_{L,d}$ denotes the special value for $\rho_{L,d}$ that makes
the dimensionless effective range parameter $r_{L,d}$ on the right-hand side
of Eq.~(\ref{ere}) equal to zero. After fine-tuning $a_{L,d}^{-1}=0$, then as
$p\rightarrow0$ the effective scattering parameter has a logarithmic
strong-coupling limit%
\begin{equation}
\alpha_{L,d}^{-1}(p)\approx-\frac{2}{\pi}p^{2}\ln\left(  -ip\bar{\rho}%
_{L,d}\right)  \rightarrow0.
\end{equation}
We note the emergence of a second relevant dimensionful parameter, $\bar{\rho
}_{L,d}$. \ In the limit $\left\vert a_{L,d}\right\vert \rightarrow\infty$,
our causality bound for $2L+d=4$ places an upper bound on $\bar{\rho}_{L,d}$,%
\begin{equation}
\bar{\rho}_{L,d}\leq\frac{R}{2}e^{\gamma}.
\end{equation}

For $2L+d\geq5$, we have%
\begin{equation}
\alpha_{L,d}^{-1}(p)=a_{L,d}^{-1}-\frac{1}{2}r_{L,d}p^{2}+\left\{
\begin{array}
[c]{c}%
O(p^{3})\text{ for }2L+d=5.\\
O(p^{4}\ln p)\text{ for }2L+d=6.\\
O(p^{4})\text{ otherwise}.
\end{array}
\right.
\end{equation}
Again we first fine-tune $a_{L,d}^{-1}=0$. \ This produces a power-law
strong-coupling limit proportional to the dimensionful parameter, $r_{L,d}$,%
\begin{equation}
\alpha_{L,d}^{-1}(p)\approx-\frac{1}{2}r_{L,d}p^{2}\rightarrow0.
\end{equation}
Since this is not scale invariant we might consider a second fine-tuning where
$r_{L,d}$ is also tuned to zero. \ However, this is not allowed by causality.
\ In the limit $\left\vert a_{L,d}\right\vert \rightarrow\infty$ our causality
bound for $2L+d\geq5$ places an upper bound on $r_{L,d},$%
\begin{equation}
r_{L,d}\leq-\frac{2\Gamma(L+\frac{d}{2}-2)\Gamma(L+\frac{d}{2}-1)}{\pi}\left(
\frac{R}{2}\right)  ^{-2L-d+4}.
\end{equation}
Since the expression on the right-hand side is a fixed negative number, the
value $r_{L,d}=0$ is not allowed.

We see that for $2L+d\geq4$ we are left with two relevant parameters which
parametrize the strong-coupling low-energy limit. \ This corresponds to two
relevant directions near a fixed point of the renormalization group, and the
universal behavior is characterized by two low-energy parameters. \ For the
case of $P$-wave neutron-alpha scattering in three dimensions, this issue was
already discussed in \cite{Bertulani:2002sz}. \ Proper renormalization of an
effective field theory for $P$-wave scattering requires the inclusion of field
operators for the scattering volume and the effective range at leading order.
\ In the renormalization group study of \cite{Barford:2002je}, the emergence
of two relevant directions around a fixed point was observed for various model potentials.

\section{Impact of Causality Bounds}

\label{sec:impact}

In the following, we illustrate the impact of the causality bounds for three
examples. We start with a spherical step potential in $d$ dimensions. This
corresponds to a purely short-range interaction and our causality bounds apply
strictly. As a second example, we consider the neutron-alpha interaction which
is characterized by resonant $P$-wave interactions. The interaction is
mediated by pion exchange which corresponds to an exponentially-bounded
interaction of $O(e^{-r/R})$ at large distances. In this case, the
results should still be accurate with only exponentially small corrections.
Finally, we consider the long-range van der Waals interaction where our
general bounds do not apply. We show how our treatment must be modified in
this case relevant to ultracold atoms.

\subsection{Spherical step potential in $d$ dimensions}

As an example of the results discussed we consider a spherical step potential
with radius $R$ and depth $V_{\text{step}}$,%
\begin{equation}
W(r,r^{\prime})=V_{\text{step}}\theta(R-r)\delta(r-r^{\prime}).
\end{equation}
We define $\kappa$ so that $\kappa^{2}=-2\mu V_{\text{step}}$ and $p^{\prime
}=\sqrt{p^{2}+\kappa^{2}}$. \ A repulsive step corresponds with positive
imaginary $\kappa$ and an attractive step corresponds with a positive real
$\kappa$. \ For the exterior region, $r\geq R,$ the wave function
$u_{L,d}^{(p)}(r)$ is given by Eq.~(\ref{u_Ld}). \ For the interior region,
$r<R$, the wave function is%
\begin{equation}
u_{L,d}^{(p)}(r)\propto\sqrt{\frac{p^{\prime}r\pi}{2}}J_{L+d/2-1}(p^{\prime
}r).
\end{equation}
Matching at the boundary $r=R$ we find%
\begin{align}
&  \cot\delta_{L,d}(p)\nonumber\\
&  =\frac{pRJ_{\alpha}(p^{\prime}R)\left[  Y_{\alpha-1}(pR)-Y_{\alpha
+1}(pR)\right]  -p^{\prime}RY_{\alpha}(pR)\left[  J_{\alpha-1}(p^{\prime
}R)-J_{\alpha+1}(p^{\prime}R)\right]  }{pRJ_{\alpha}(p^{\prime}R)\left[
J_{\alpha-1}(pR)-J_{\alpha+1}(pR)\right]  -p^{\prime}RJ_{\alpha}(pR)\left[
J_{\alpha-1}(p^{\prime}R)-J_{\alpha+1}(p^{\prime}R)\right]  },
\label{cot_step}%
\end{align}
where $\alpha=L+d/2-1$.

\begin{figure}[ptb]
\begin{center}
\includegraphics[
height=3.0225in,
width=3.8017in
]{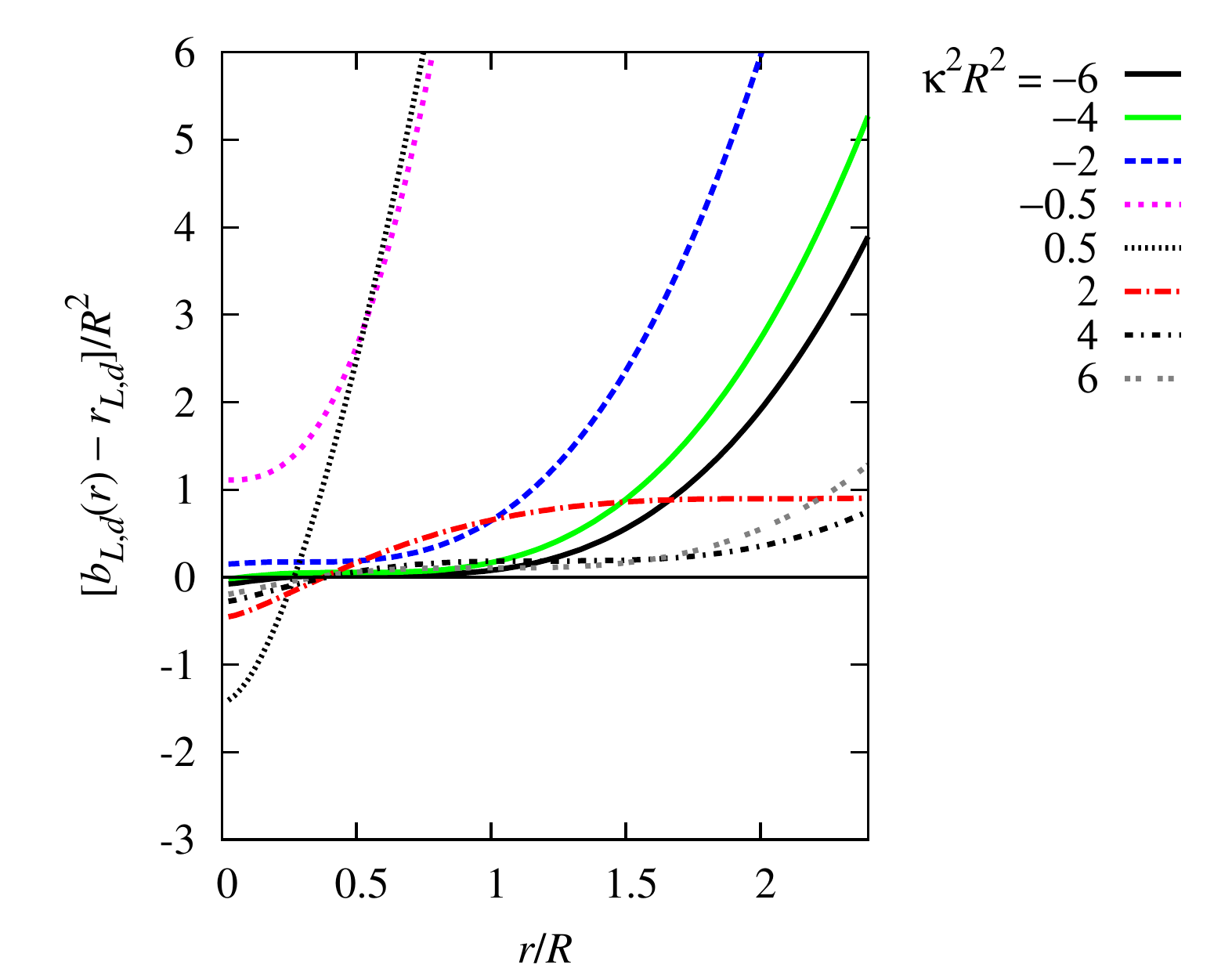}
\end{center}
\caption{Plot of $\left[  b_{L,d}(r)-r_{L,d}\right]  /R^{2}$ as a function of
$r/R$ for spherical step potentials for $2L+d=2$.}%
\label{d2l_2}%
\end{figure}

\begin{figure}[ptbptb]
\begin{center}
\includegraphics[
height=3.0225in,
width=3.8017in
]{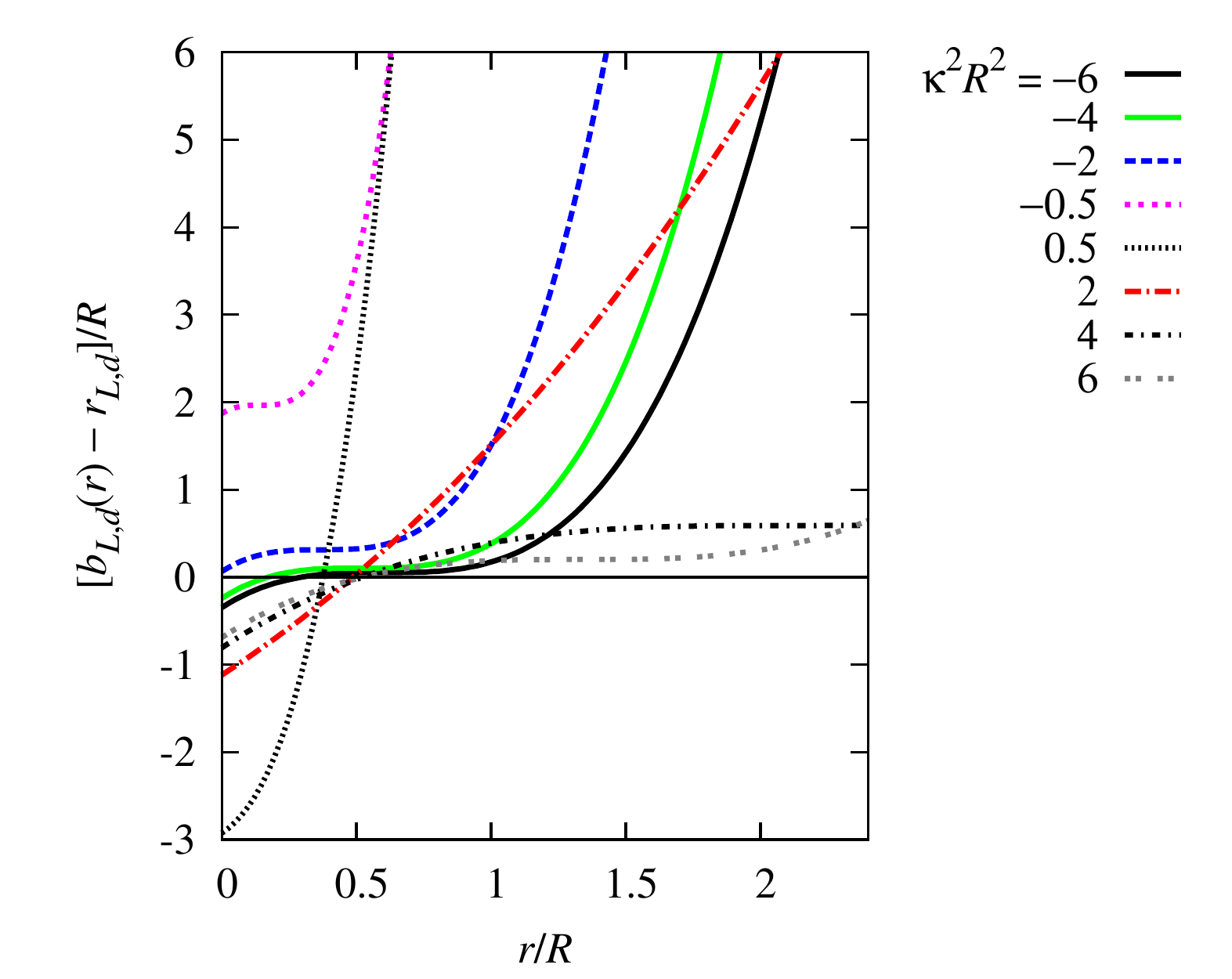}
\end{center}
\caption{Plot of $\left[  b_{L,d}(r)-r_{L,d}\right]  /R$ as a function of
$r/R$ for spherical step potentials for $2L+d=3$.}%
\label{d2l_3}%
\end{figure}

We use Eq.~(\ref{cot_step}) to generate the effective range expansion for%
\begin{equation}
p^{2L+d-2}\left[  \cot\delta_{L,d}(p)-\delta_{(d\operatorname{mod}2),0}%
\frac{2}{\pi}\ln\left(  p\rho_{L,d}\right)  \right]  .
\end{equation}
For $2L+d=1,$ the lowest two coefficients in the effective range expansion are%
\begin{equation}
a_{L,d}^{-1}R^{-1}=-\frac{\kappa R+\cot\left(  \kappa R\right)  }{\kappa R},
\label{step_ainv_1}%
\end{equation}%
\begin{equation}
r_{L,d}R^{-3}=\frac{\kappa R\left(  2\kappa^{2}R^{2}-3\right)  +3(2\kappa
^{2}R^{2}-1)\cot\left(  \kappa R\right)  +3\kappa R\cot^{2}\left(  \kappa
R\right)  }{3\kappa^{3}R^{3}}. \label{step_effr_1}%
\end{equation}
In each case we multiply by powers of $R$ to render the quantity
dimensionless. \ For $2L+d=2$, we use the convention $\rho_{L,d}=\frac{R}{2}$.
\ In this case the dimensionless coefficients are%
\begin{equation}
a_{L,d}^{-1}=-\frac{2}{\pi}\left[  \gamma+\frac{J_{0}(\kappa R)}{\kappa
RJ_{1}(\kappa R)}\right]  , \label{step_ainv_2}%
\end{equation}%
\begin{equation}
r_{L,d}R^{-2}=\frac{1}{\pi\kappa^{2}R^{2}}\left[  \kappa^{2}R^{2}-2+2\kappa
R\frac{J_{0}(\kappa R)}{J_{1}(\kappa R)}\right]  . \label{step_effr_2}%
\end{equation}
For $2L+d=3,$%
\begin{equation}
a_{L,d}^{-1}R=\frac{\kappa R\cos\left(  \kappa R\right)  }{\kappa R\cos\left(
\kappa R\right)  -\sin\left(  \kappa R\right)  }, \label{step_ainv_3}%
\end{equation}%
\begin{equation}
r_{L,d}R^{-1}=\frac{2\kappa^{3}R^{3}+2\kappa R\left(  \kappa^{2}%
R^{2}-3\right)  \cos\left(  2\kappa R\right)  +3(-2\kappa^{2}R^{2}%
+1)\sin\left(  2\kappa R\right)  }{6\kappa R\left[  -\kappa R\cos\left(
\kappa R\right)  +\sin\left(  \kappa R\right)  \right]  ^{2}}.
\label{step_effr_3}%
\end{equation}
For $2L+d=4$, we again use the convention $\rho_{L,d}=\frac{R}{2}$,%
\begin{equation}
a_{L,d}^{-1}R^{2}=-\frac{4}{\pi}\frac{J_{0}\left(  \kappa R\right)  }%
{J_{2}\left(  \kappa R\right)  }, \label{step_ainv_4}%
\end{equation}%
\begin{align}
&  r_{L,d}=\left\{  8\left(  2\gamma-1\right)  \kappa RJ_{1}^{2}\left(  \kappa
R\right)  +\kappa RJ_{0}^{2}\left(  \kappa R\right)  \left[  \left(
4\gamma-3\right)  \kappa^{2}R^{2}-4\right]  +4\kappa RJ_{0}\left(  \kappa
R\right)  J_{2}\left(  \kappa R\right)  \right. \nonumber\\
&  \left.  +8J_{0}\left(  \kappa R\right)  J_{1}\left(  \kappa R\right)
\left[  \left(  -2\gamma+1\right)  \kappa^{2}R^{2}+1\right]  \right\}
\times\frac{1}{\pi\kappa R\left[  \kappa RJ_{0}\left(  \kappa R\right)
-2J_{1}\left(  \kappa R\right)  \right]  ^{2}}. \label{step_effr_4}%
\end{align}
For $2L+d=5,$%
\begin{equation}
a_{L,d}^{-1}R^{3}=\frac{3\kappa^{2}R^{2}}{-3+\kappa^{2}R^{2}+3\kappa
R\cot\left(  \kappa R\right)  }, \label{step_ainv_5}%
\end{equation}%
\begin{align}
&  r_{L,d}R=\left\{  -18\kappa R\left(  \kappa^{2}R^{2}+5\right)  +18\kappa
R\left(  \kappa^{2}R^{2}-10\right)  \cos\left(  2\kappa R\right)  \right.
\nonumber\\
&  \left.  +45\left(  -2\kappa^{2}R^{2}+3\right)  \sin\left(  2\kappa
R\right)  \right\}  \times\frac{\kappa R}{10\left[  3\kappa R\cos\left(
\kappa R\right)  +\left(  \kappa^{2}R^{2}-3\right)  \sin\left(  \kappa
R\right)  \right]  ^{2}}. \label{step_effr_5}%
\end{align}
In Figs.~\ref{d2l_2} and \ref{d2l_3}, we plot the function $\left[
b_{L,d}(r)-r_{L,d}\right]  R^{2L+d-4}$ as a function of $r/R$ for the sample
values $2L+d=2,\,3$. \ We see that the function is non-negative for $r/R\geq
1$, as required by causality. \ As $\kappa^{2}R^{2}\rightarrow-\infty$, the
potential becomes a hard spherical barrier with the wave function
$u_{L,d}^{(p)}(r)$ vanishing in the interior region, $r\leq R$. \ Since%
\begin{equation}
\int_{0}^{R}dr^{\prime}\left[  u_{L,d}^{(0)}(r^{\prime})\right]
^{2}\rightarrow0^{+},
\end{equation}
the causality bound is saturated in the hard barrier limit for $r=R$,%
\begin{equation}
b_{L,d}(R)-r_{L,d}\rightarrow0^{+}.
\end{equation}
For the other values of $2L+d$ not shown in Figs.~\ref{d2l_2} and \ref{d2l_3},
the behavior is qualitatively the same.


\subsection{Alpha-neutron scattering}

Our results are exact only for the case where the interaction vanishes for
$r\geq R$. \ For exponentially-bounded interactions of $O(e^{-r/R})$
at large distances, the results should still be accurate with only
exponentially small corrections. \ For an exponentially-bounded but otherwise
unknown interaction, the non-negativity condition for $b_{L,d}(r)-r_{L,d}$ can
be used to determine the minimum value for $R$ consistent with causality.
\ One example of an exponentially-bounded interaction is the three-dimensional
scattering of an alpha particle and neutron. \ We consider $S$-wave and
$P$-wave alpha-neutron scattering.

\begin{figure}[ptb]
\begin{center}
\includegraphics[
height=2.8236in,
width=2.9628in
]{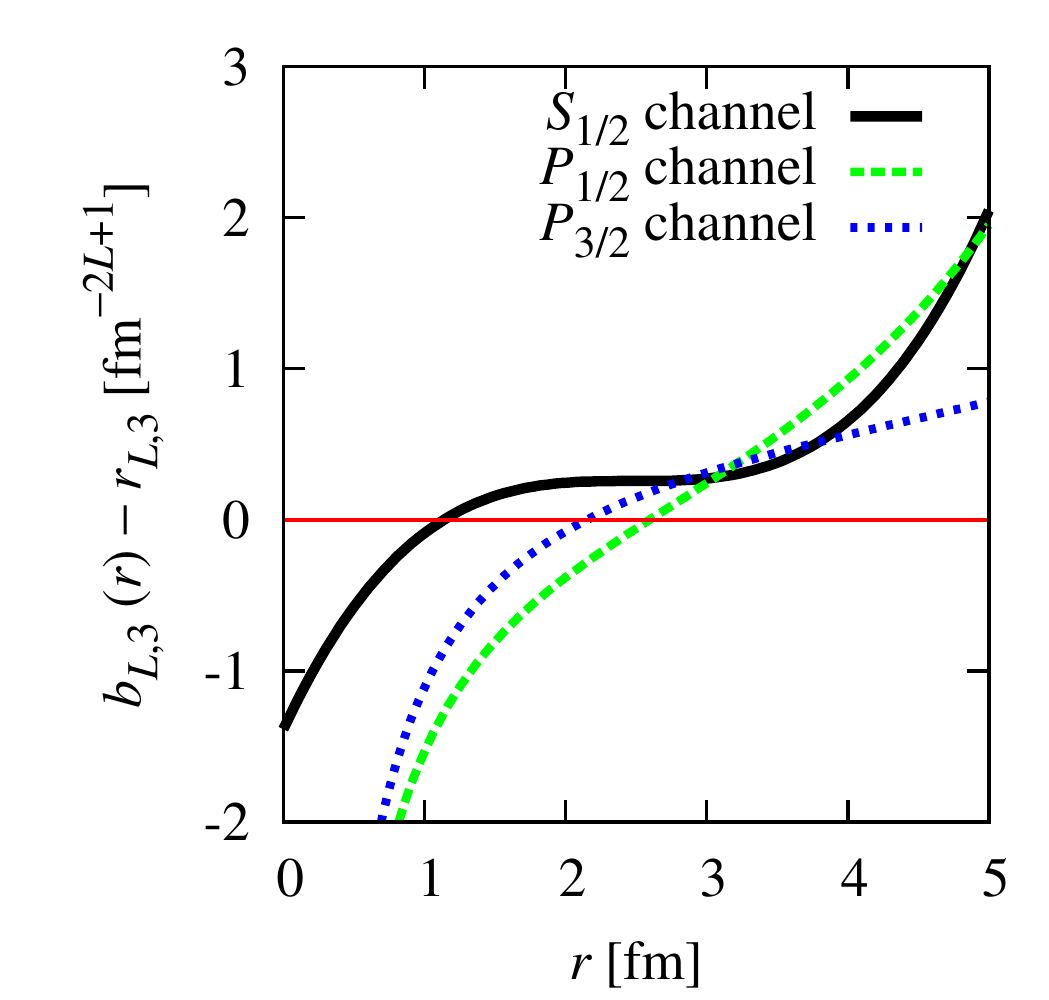}
\end{center}
\caption{Plot of $b_{L,3}(r)-r_{L,3}$ as a function of $r$ for alpha-neutron
scattering in the $S_{1/2}$, $P_{1/2}$, and $P_{3/2}$ channels.}%
\label{alpha_neutron}%
\end{figure}

For $S$-wave scattering,%
\begin{equation}
b_{0,3}(r)=2r-\frac{2r^{2}}{a_{0,3}}+\frac{2r^{3}}{3a_{0,3}^{2}},
\end{equation}
and for $P$-wave scattering,%
\begin{equation}
b_{1,3}(r)=-\frac{2}{r}-\frac{2r^{2}}{3a_{1,3}}+\frac{2r^{5}}{45a_{1,3}^{2}}.
\end{equation}
In Fig.~\ref{alpha_neutron} we plot $b_{L,3}(r)-r_{L,3}$ for the $S_{1/2}$,
$P_{1/2}$, and $P_{3/2}$ channels. \ A qualitatively similar plot was
introduced for nucleon-nucleon scattering in the $S$-wave spin-singlet channel
\cite{Scaldeferri:1996nx}. \ We use the values $a_{0,3}=2.464(4)$ fm and
$r_{0,3}=1.39(4)$ fm for $S_{1/2}$; $a_{1,3}=-13.82(7)$ fm$^{3}$ and
$r_{1,3}=-0.42(2)$ fm$^{\text{-1}}$ for $P_{1/2}$; and $a_{1,3}=-62.951(3)$
fm$^{3}$ and $r_{1,3}=-0.882(1)$ fm$^{\text{-1}}$ for $P_{3/2}$
\cite{Arndt:1973a}. \ The non-negativity condition for $b_{L,3}(r)-r_{L,3}$
gives $R\geq1.1$ fm for $S_{1/2}$, $R\geq2.6$ fm for $P_{1/2}$, and $R\geq2.1$
fm for $P_{3/2}$. \ For comparison, the alpha root-mean-square radius and pion
Compton wavelength are both about $1.5$ fm. \ Since the minimum values for $R$
are not small when compared with these, some caution is required when choosing
the cutoff scale for an effective theory of alpha-neutron interactions.

We can use the results for the spherical step potential to reproduce the
scattering parameter and effective range parameter for alpha-neutron
scattering. \ For the $S$-wave, $r_{0,3}a_{0,3}^{-1}=0.564$. \ This gives
$\kappa^{2}R^{2}=-5.56$. \ In turn this implies $r_{0,3}/R=0.329$ and
therefore $R=4.23$ fm. \ For the $P_{1/2}$-channel, $r_{1,3}^{-3}a_{1,3}%
^{-1}=0.977.$ \ This gives $\kappa^{2}R^{2}=5.034$, $r_{1,3}R=-1.67$, and so
$R=3.97$ fm. \ For the $P_{3/2}$-channel, $r_{1,3}^{-3}a_{1,3}^{-1}=0.0232$.
\ This gives $\kappa^{2}R^{2}=8.90$, $r_{1,3}R=-2.84$, and $R=3.22$ fm.

\subsection{Van der Waals interaction}

The physics of long-range interactions must be treated separately since each
long-range behavior determines its own low-energy universality class. \ For
cold alkali atoms our analysis must be modified to take into account
long-range van der Waals interactions of the type%
\begin{equation}
W(r,r^{\prime})=-C_{6}r^{-6}\delta(r-r^{\prime})
\end{equation}
for $r,r^{\prime}\geq R$. \ It is convenient to reexpress $C_{6}$ in terms of
the length scale $\beta_{6}=(2\mu C_{6})^{1/4}$. \ In the following, we set
$d=3$ and drop the $d$ subscript. \ Instead of free Bessel functions,
scattering states should be compared with exact solutions of the attractive
$r^{-6}$ potential \cite{Gao:1998a, Gao:1998b}. \ The effect of the
interactions for $r<R$ are described by a finite-range $K$-matrix $K_{L}%
(p^{2})$ that is analytic in $p^{2}$ \cite{Gao:2009a},%
\begin{equation}
K_{L}(p^{2})=\sum_{n=0,1,\cdots}K_{L}^{(2n)}p^{2n}.
\end{equation}
When phase shifts are measured relative to free spherical Bessel functions,
the effective range expansion is no longer analytic in $p^{2}$. \ For $L=0$,
the leading non-analytic term is proportional to $p^{3}$,%
\begin{align}
&  p\cot\delta_{0}(p)\nonumber\\
&  =-\frac{\left[  \Gamma(1/4)\right]  ^{2}K_{0}^{(0)}}{2\pi\beta_{6}\left[
K_{0}^{(0)}-1\right]  }+\frac{\left[  \Gamma(1/4)\right]  ^{2}}{6\pi\beta_{6}%
}\frac{\beta_{6}^{2}\left(  K_{0}^{(0)}\right)  ^{2}+\beta_{6}^{2}%
+3K_{0}^{(2)}}{\left[  K_{0}^{(0)}-1\right]  ^{2}}p^{2}\nonumber\\
&  -\frac{\left[  \Gamma(1/4)\right]  ^{4}}{60\pi}\frac{\beta_{6}^{2}\left(
K_{0}^{(0)}\right)  ^{2}}{\left[  K_{0}^{(0)}-1\right]  ^{2}}p^{3}+O(p^{4}\ln
p).
\end{align}
For $L=1$ the leading non-analytic term is proportional to $p^{1}$,%
\begin{align}
&  p^{3}\cot\delta_{1}(p)\nonumber\\
&  =\frac{18\left[  \Gamma(3/4)\right]  ^{2}K_{1}^{(0)}}{\pi\beta_{6}%
^{3}\left[  K_{1}^{(0)}+1\right]  }+\frac{324\left[  \Gamma(3/4)\right]
^{4}\left(  K_{1}^{(0)}\right)  ^{2}}{35\pi\beta_{6}^{2}\left[  K_{1}%
^{(0)}+1\right]  ^{2}}p\nonumber\\
&  +\frac{-4410\left[  \Gamma(3/4)\right]  ^{2}\left[  K_{1}^{(0)}+1\right]
\left\{  \beta_{6}^{2}\left[  \left(  K_{1}^{(0)}\right)  ^{2}+1\right]
-5K_{1}^{(2)}\right\}  +5832\left[  \Gamma(3/4)\right]  ^{6}\beta_{6}%
^{2}\left(  K_{1}^{(0)}\right)  ^{3}}{1225\pi\beta_{6}^{3}\left[  K_{1}%
^{(0)}+1\right]  ^{3}}p^{2}\nonumber\\
&  +O(p^{3}).
\end{align}
This term voids the standard definition of the effective range parameter for
$P$-waves.

However, one can still obtain useful information. \ The zero-energy resonance
limit is reached by tuning the lowest-order $K$-matrix coefficient
$K_{L}^{(0)}$ to zero. \ In this limit the leading non-analytic terms in the
effective range expansion vanishes, and one can define an effective range
parameter for both $S$- and $P$-waves \cite{Gao:1998b,Flambaum:1999zza},%
\begin{equation}
r_{0}=\frac{\left[  \Gamma(1/4)\right]  ^{2}\left[  \beta_{6}^{2}+3K_{0}%
^{(2)}\right]  }{3\pi\beta_{6}},
\end{equation}%
\begin{equation}
r_{1}=\frac{-36\left[  \Gamma(3/4)\right]  ^{2}\left[  \beta_{6}^{2}%
-5K_{1}^{(2)}\right]  }{5\pi\beta_{6}^{3}}.
\end{equation}
For the case of single-channel scattering of alkali atoms, the coefficients
$K_{L}^{(2)}$ are negligible compared with $\beta_{6}^{2}$. \ This is also
true for some multi-channel Feshbach resonance systems \cite{Hanna:2009a}.
\ In these cases we observe that the upper bounds for $r_{L}$ in
Eq.~(\ref{2Ld_odd}) are satisfied for $L=0$ and $L=1$ when we naively take
$R\sim\beta_{6}$. \ In general, there may be multi-channel systems where the
coefficients $K_{L}^{(2)}$ cannot be neglected. \ Nevertheless, the
coefficients $K_{L}^{(2)}$ should satisfy causality bounds similar to those
derived here for the effective range parameter.

\section{Summary and Outlook}

\label{sec:sum}

In this paper, we have addressed the question of universality\ and the
constraints of causality on quantum scattering processes for arbitrary
dimension $d$ and arbitrary angular momentum $L$. We have derived the
Wronskian identity for two solutions of the radial Schr\"odinger equation with
different energy and used this identity to generalize the causality bound on
the effective range to arbitrary $d$ and $L$. Moreover, we have derived a
general causality bound for energies away from threshold.

For finite-range interactions, we have shown that causal wave propagation can
have significant consequences for low-energy universality and scale
invariance. \ For $2L+d\geq4$, two relevant low-energy parameters are required
in the strong-coupling low-energy limit. \ In the language of the
renormalization group, this corresponds to two relevant directions in the
vicinity of a fixed point. \ In particular, we confirm earlier findings for 
three-dimensional $P$-wave scattering \cite{Bertulani:2002sz} based on
renormalization arguments and for higher partial waves in general
\cite{Barford:2002je} in the framework of the renormalization group.

In the low-energy limit, the hierarchy of terms in the effective range
expansion depends on the value of $2L+d$ (cf.~Fig.~\ref{staircase}). In
particular, the effective range parameter is as important at low energies as
the unitarity contribution for $2L+d=4$ and becomes more important for $2L+d
\geq5$. Our results imply that the scale-invariant unitarity limit can not be
reached in this case because the causality bound prevents the effective range
from being tuned to zero. This has important consequences for the universal
properties of systems with $P$-wave and higher partial wave interactions.
Causality also constrains the wave functions and the probability to find
particles at large spatial separation \cite{Hammer:2009zh}. This issue will be
discussed in a separate publication.

We stress that our results strictly apply only to energy-independent
interactions and single-channel systems. For energy-dependent interactions it
is possible to generate any energy dependence for the elastic phase shifts even
when the interaction $W(r,r^{\prime};E)$ vanishes beyond some finite radius
$R$ for all $E$. \ Under these more general conditions, there are no longer
any Wigner bounds and the constraints of causality seem to disappear.
\ However, it is misleading to regard interactions of this more general type
as having finite range. \ As noted in the introduction, the scattering time
delay is given by the energy derivative of the phase shift. \ The energy
dependence of the interaction can by itself generate large negative time
delays and thereby reproduce the scattering of long-range interactions. In
this sense the range of the interaction as observed in scattering is set by
the dependence of $W(r,r^{\prime};E)$ on the radial coordinates $r,r^{\prime}$
as well as the energy $E$. \ In this case our bound can be viewed as an
estimate for the minimum value of this interaction range. For coupled-channel
dynamics without partial wave mixing the analysis can proceed by first
integrating out higher-energy contributions to produce a single-channel
effective interaction. \ In order to satisfy our condition of
energy-independent interactions, this should be carried out using a technique
such as the method of unitary transformation described in
Ref.~\cite{Fukuda:1954a, Okubo:1954zz, Epelbaum:1998ka}.

Our analysis concerns only the question of universality in two-body
scattering. \ Universality for higher few-body systems requires a detailed
analysis for each system under consideration. For resonant $S$-wave
interactions, the question of universality has already been explored for three
and more particles \cite{Braaten:2004a}. In two dimensions, the properties of
$N$-boson droplets are universal for $N$ large, but below some critical value
\cite{Hammer:2004as,Blume:2005zz,Lee:2005nm}. In three dimensions, the Efimov
effect generates a universal spectrum of shallow three-body bound states
\cite{Efimov:1970zz} and two universal four-body states are attached to each
three-body Efimov state \cite{Hammer:2006ct,vStech08}. Effective field theory
and renormalization group methods may provide a useful starting point to
extend these studies to more particles and higher angular momentum
\cite{Bedaque:1998kg,Barford:2004fz,Moroz:2008fy,Krippa:2009vu}. Our results
help to clarify some of the conceptual and calculational issues relevant to
few-body systems for general dimension and angular momentum and their
simulation using short-range interactions.

\section*{Acknowledgements}

We are grateful for discussions with D. Phillips and T. Sch\"{a}fer. \ This
research was supported in part by the DFG through SFB/TR 16 \textquotedblleft
Subnuclear structure of matter,\textquotedblright\ the BMBF under contract
06BN9006, and the US Department of Energy under DE-FG02-03ER41260.

\appendix

\section{Expressions for $b_{L,d}(r)$}

\label{app:b_Ld}

In this Appendix, we check explicitly that the identities Eqs.~(\ref{eq:idW1},
\ref{eq:idW2}) hold for all combinations of $d$ and $L$ and derive explicit
expressions for $b_{L,d}(r)$. Moreover, our conventions for the Riccati-Bessel
and Bessel functions are given in Eqs.~(\ref{eq:BesselS}, \ref{eq:BesselC},
\ref{eq:BesselJ}, \ref{eq:BesselY}).

\subsection{Positive odd integer $2L+d$}


For positive odd integer $2L+d$ it is convenient to use the second line of
Eq.~(\ref{u_Ld}). \ For $r\geq R$,%
\begin{equation}
u_{L,d}^{(p)}(r)=p^{L+d/2-3/2}\left[  \cot\delta_{L,d}(p)S_{L+d/2-3/2}%
(pr)+C_{L+d/2-3/2}(pr)\right]  .
\end{equation}
Writing this in the form dictated in Eq.~(\ref{standard_form}), we have%
\begin{equation}
s(p,r)=p^{-L-d/2+1/2}S_{L+d/2-3/2}(pr),
\end{equation}%
\begin{equation}
c(p,r)=p^{L+d/2-3/2}C_{L+d/2-3/2}(pr).
\end{equation}
When $2L+d$ is a positive odd integer, $L+d/2-3/2$ is an integer greater than
or equal to $-1$. \ For integer $n$, the Riccati-Bessel functions are given by%
\begin{equation}
S_{n}(x)=\sqrt{\pi}x^{n+1}\sum_{m=0}^{\infty}\frac{i^{2m}2^{-2m-n-1}}%
{\Gamma(m+1)\Gamma(m+n+\frac{3}{2})}x^{2m} \label{eq:BesselS}%
\end{equation}%
\begin{equation}
C_{n}(x)=\frac{1}{\sqrt{\pi}}x^{-n}\Gamma(-n+\frac{1}{2})\Gamma(n+\frac{1}%
{2})\sum_{m=0}^{\infty}\frac{i^{2m}2^{-2m+n}}{\Gamma(m+1)\Gamma(m-n+\frac
{1}{2})}x^{2m}. \label{eq:BesselC}%
\end{equation}
The first two series coefficients for $s(p,r)$ and $c(p,r)$ in powers of
$p^{2}$ are%
\begin{equation}
s_{0}(r)=\frac{\sqrt{\pi}}{\Gamma(L+\frac{d}{2})}\left(  \frac{r}{2}\right)
^{L+d/2-1/2},\text{\qquad}s_{2}(r)=-\frac{\sqrt{\pi}}{\Gamma(L+\frac{d}{2}%
+1)}\left(  \frac{r}{2}\right)  ^{L+d/2+3/2},
\end{equation}%
\begin{equation}
c_{0}(r)=\frac{\Gamma(L+\frac{d}{2}-1)}{\sqrt{\pi}}\left(  \frac{r}{2}\right)
^{-L-d/2+3/2},\text{\qquad}c_{2}(r)=\frac{\Gamma(L+\frac{d}{2}-2)}{\sqrt{\pi}%
}\left(  \frac{r}{2}\right)  ^{-L-d/2+7/2}\text{.}%
\end{equation}
The corresponding Wronskians are%
\begin{equation}
W[s_{0},c_{0}](r)=-1,
\end{equation}%
\begin{equation}
W[s_{2},s_{0}](r)=\frac{\pi}{\Gamma(L+\frac{d}{2})\Gamma(L+\frac{d}{2}%
+1)}\left(  \frac{r}{2}\right)  ^{2L+d},
\end{equation}%
\begin{equation}
W[s_{2},c_{0}](r)=W[c_{2},s_{0}](r)=\frac{r^{2}}{2(2L+d-2)},
\end{equation}%
\begin{equation}
W[c_{2},c_{0}](r)=-\frac{\Gamma(L+\frac{d}{2}-2)\Gamma(L+\frac{d}{2}-1)}{\pi
}\left(  \frac{r}{2}\right)  ^{-2L-d+4}.
\end{equation}
We conclude that for $r\geq R$,%
\begin{align}
b_{L,d}(r)  &  =-\frac{2\Gamma(L+\frac{d}{2}-2)\Gamma(L+\frac{d}{2}-1)}{\pi
}\left(  \frac{r}{2}\right)  ^{-2L-d+4}\nonumber\\
&  -\frac{4}{L+\frac{d}{2}-1}\frac{1}{a_{L,d}}\left(  \frac{r}{2}\right)
^{2}\nonumber\\
&  +\frac{2\pi}{\Gamma(L+\frac{d}{2})\Gamma(L+\frac{d}{2}+1)}\frac{1}%
{a_{L,d}^{2}}\left(  \frac{r}{2}\right)  ^{2L+d}. \label{b_Ld_odd}%
\end{align}
\bigskip

\subsection{Positive even integer $2L+d$}


For positive even integer $2L+d$ it is convenient to use the first line of
Eq.~(\ref{u_Ld}). \ For $r\geq R$,%
\begin{equation}
u_{L,d}^{(p)}(r)=\sqrt{\frac{pr\pi}{2}}p^{L+d/2-3/2}\left[  \cot\delta
_{L,d}(p)J_{L+d/2-1}(pr)-Y_{L+d/2-1}(pr)\right]  .
\end{equation}
When $2L+d$ is a positive even integer, $L+d/2-1$ is a non-negative integer.
\ For non-negative integer $n$ we have%
\begin{equation}
J_{n}(x)=\sum_{m=0}^{\infty}\frac{(-1)^{m}}{\Gamma(m+1)\Gamma(m+n+1)}\left(
\frac{x}{2}\right)  ^{2m+n}, \label{eq:BesselJ}%
\end{equation}
and
\begin{align}
Y_{n}(x)  &  =\frac{2}{\pi}\left(  \ln\frac{x}{2}+\gamma\right)
J_{n}(x)-\frac{1}{\pi}\sum_{k=0}^{n-1}\frac{\Gamma(n-k)}{\Gamma(k+1)}\left(
\frac{x}{2}\right)  ^{2k-n}\nonumber\\
&  -\frac{1}{\pi}\sum_{m=0}^{\infty}\frac{(-1)^{m}\left(  H_{m}+H_{n+m}%
\right)  }{\Gamma(m+1)\Gamma(m+n+1)}\left(  \frac{x}{2}\right)  ^{2m+n}
\label{eq:BesselY}%
\end{align}
$\gamma$ is the Euler-Mascheroni constant, and $H_{k}$ is the $k^{\text{th}}$
harmonic number,%
\begin{equation}
H_{k}=%
{\displaystyle\sum\limits_{m=1}^{k}}
\frac{1}{m}.
\end{equation}
We can write $u_{L,d}^{(p)}(r)$ in the form dictated in
Eq.~(\ref{standard_form}), if we let%
\begin{equation}
s(p,r)=\sqrt{\frac{pr\pi}{2}}p^{-L-d/2+1/2}J_{L+d/2-1}(pr)
\end{equation}%
\begin{equation}
c(p,r)=-\sqrt{\frac{pr\pi}{2}}p^{L+d/2-3/2}\left[  Y_{L+d/2-1}(pr)-\frac
{2}{\pi}\ln\left(  p\rho_{L,d}\right)  J_{L+d/2-1}(pr)\right]  .
\end{equation}
We now consider each of the possible cases for positive even integer $2L+d$.

\subsubsection{Case 2$L+d=2$}

When $2L+d=2$ we have%
\begin{equation}
s(p,r)=\sqrt{\frac{r\pi}{2}}J_{0}(pr)
\end{equation}%
\begin{equation}
c(p,r)=-\sqrt{\frac{r\pi}{2}}\left[  Y_{0}(pr)-\frac{2}{\pi}\ln\left(
p\rho_{L,d}\right)  J_{0}(pr)\right]  .
\end{equation}
Both functions are analytic in $p^{2}$. \ The first two series coefficients
are%
\begin{equation}
s_{0}(r)=\sqrt{\pi}\left(  \frac{r}{2}\right)  ^{1/2},\qquad s_{2}%
(r)=-\sqrt{\pi}\left(  \frac{r}{2}\right)  ^{5/2}%
\end{equation}%
\begin{equation}
c_{0}(r)=-\frac{2}{\sqrt{\pi}}\left(  \frac{r}{2}\right)  ^{1/2}\left[
\ln\left(  \frac{r}{2\rho_{L,d}}\right)  +\gamma\right]  ,\qquad
c_{2}(r)=\frac{2}{\sqrt{\pi}}\left(  \frac{r}{2}\right)  ^{5/2}\left[
\ln\left(  \frac{r}{2\rho_{L,d}}\right)  +\gamma-1\right]  .
\end{equation}
The corresponding Wronskians are%
\begin{equation}
W[s_{0},c_{0}](r)=-1,
\end{equation}%
\begin{equation}
W[s_{2},s_{0}](r)=\pi\left(  \frac{r}{2}\right)  ^{2},
\end{equation}%
\begin{equation}
W[s_{2},c_{0}](r)=W[c_{2},s_{0}](r)=-2\left[  \ln\left(  \frac{r}{2\rho_{L,d}%
}\right)  +\gamma-\frac{1}{2}\right]  \left(  \frac{r}{2}\right)  ^{2},
\end{equation}%
\begin{equation}
W[c_{2},c_{0}](r)=\frac{4}{\pi}\left\{  \left[  \ln\left(  \frac{r}%
{2\rho_{L,d}}\right)  +\gamma-\frac{1}{2}\right]  ^{2}+\frac{1}{4}\right\}
\left(  \frac{r}{2}\right)  ^{2}.
\end{equation}
The function $b_{L,d}(r)$ for this case is%
\begin{equation}
b_{L,d}(r)=\frac{2r^{2}}{\pi}\left\{  \left[  \ln\left(  \frac{r}{2\rho_{L,d}%
}\right)  +\gamma-\frac{1}{2}+\frac{\pi}{2a_{L,d}}\right]  ^{2}+\frac{1}%
{4}\right\}  . \label{b_Ld_2}%
\end{equation}

\subsubsection{Case 2$L+d=4$}

For $2L+d=4$,%
\begin{equation}
s(p,r)=\sqrt{\frac{\pi r}{2}}p^{-1}J_{1}(pr),
\end{equation}%
\begin{equation}
c(p,r)=-\sqrt{\frac{\pi r}{2}}p\left[  Y_{1}(pr)-\frac{2}{\pi}\ln\left(
p\rho_{L,d}\right)  J_{1}(pr)\right]  .
\end{equation}
The series coefficients are%
\begin{equation}
s_{0}(r)=\sqrt{\pi}\left(  \frac{r}{2}\right)  ^{3/2},\qquad s_{2}%
(r)=-\frac{\sqrt{\pi}}{2}\left(  \frac{r}{2}\right)  ^{7/2},
\end{equation}%
\begin{equation}
c_{0}(r)=\frac{1}{\sqrt{\pi}}\left(  \frac{r}{2}\right)  ^{-1/2},\qquad
c_{2}(r)=-\frac{1}{\sqrt{\pi}}\left(  \frac{r}{2}\right)  ^{3/2}\left[
2\ln\left(  \frac{r}{2\rho_{L,d}}\right)  +2\gamma-1\right]  ,
\end{equation}
and the Wronskians are%
\begin{equation}
W[s_{0},c_{0}](r)=-1,
\end{equation}%
\begin{equation}
W[s_{2},s_{0}](r)=\frac{\pi}{2}\left(  \frac{r}{2}\right)  ^{4},
\end{equation}%
\begin{equation}
W[s_{2},c_{0}](r)=W[c_{2},s_{0}](r)=\left(  \frac{r}{2}\right)  ^{2},
\end{equation}%
\begin{equation}
W[c_{2},c_{0}](r)=\frac{2}{\pi}\left[  \ln\left(  \frac{r}{2\rho_{L,d}%
}\right)  +\gamma\right]  .
\end{equation}
The function $b_{L,d}(r)$ is%
\begin{equation}
b_{L,d}(r)=\frac{4}{\pi}\left[  \ln\left(  \frac{r}{2\rho_{L,d}}\right)
+\gamma\right]  -\frac{4}{a_{L,d}}\left(  \frac{r}{2}\right)  ^{2}+\frac{\pi
}{a_{L,d}^{2}}\left(  \frac{r}{2}\right)  ^{4}\text{.} \label{b_Ld_4}%
\end{equation}

\subsubsection{Case 2$L+d\geq6$}

The last case we consider is when $2L+d$ is an even integer greater than or
equal to $6$. \ Here we have%
\begin{equation}
s(p,r)=\sqrt{\frac{r\pi}{2}}p^{-L-d/2+1}J_{L+d/2-1}(pr)
\end{equation}%
\begin{equation}
c(p,r)=-\sqrt{\frac{r\pi}{2}}p^{L+d/2-1}\left[  Y_{L+d/2-1}(pr)-\frac{2}{\pi
}\ln\left(  p\rho_{L,d}\right)  J_{L+d/2-1}(pr)\right]  .
\end{equation}
The first two series coefficients have exactly the same form as in the case
for positive odd integer $2L+d$,%
\begin{equation}
s_{0}(r)=\frac{\sqrt{\pi}}{\Gamma(L+\frac{d}{2})}\left(  \frac{r}{2}\right)
^{L+d/2-1/2},\text{\qquad}s_{2}(r)=-\frac{\sqrt{\pi}}{\Gamma(L+\frac{d}{2}%
+1)}\left(  \frac{r}{2}\right)  ^{L+d/2+3/2},
\end{equation}%
\begin{equation}
c_{0}(r)=\frac{\Gamma(L+\frac{d}{2}-1)}{\sqrt{\pi}}\left(  \frac{r}{2}\right)
^{-L-d/2+3/2},\text{\qquad}c_{2}(r)=\frac{\Gamma(L+\frac{d}{2}-2)}{\sqrt{\pi}%
}\left(  \frac{r}{2}\right)  ^{-L-d/2+7/2}\text{.}%
\end{equation}
We conclude the same result for $b_{L,d}(r)$ as for odd $2L+d$ as written in
Eq.~(\ref{b_Ld_odd}).

\section{Equivalence with Wigner's bound}

\label{app:equiv}

In this Appendix, we demonstrate the equivalence of our causality bound with
Wigner's original bound on the energy derivative of the phase
shift~\cite{Wigner:1955a}.

Let $I_{A}(r)$ be a free incoming radial wave for momentum $p_{A}$,%
\begin{equation}
I_{A}(r)=\sqrt{\frac{p_{A}r\pi}{2}}p_{A}^{L+d/2-3/2}\left[  -i\cdot
J_{L+d/2-1}(p_{A}r)-Y_{L+d/2-1}(p_{A}r)\right]  .
\end{equation}
We are using the same phase convention as Wigner but a different
normalization. \ From Abel's differential equation identity, the Wronskian of
$I_{A}$ and $I_{A}^{\ast}$ is independent of $r$. \ For our chosen
normalization of the incoming wave,%
\begin{equation}
I_{A}(r)I_{A}^{\prime\ast}(r)-I_{A}^{\ast}(r)I_{A}^{\prime}(r)=2ip_{A}%
^{2L+d-2}. \label{I_Wronskian}%
\end{equation}
For $r\geq R$,%
\begin{equation}
u_{A}\left(  r\right)  =\frac{ie^{-i\delta_{A}}}{2\sin\delta_{L}(p)}\left[
I_{A}(r)-e^{2i\delta_{A}}I_{A}^{\ast}(r)\right]  .
\end{equation}
We define $\alpha_{A}(r)$ as the reciprocal logarithmic derivative of
$u_{A}\left(  r\right)  $,%
\begin{equation}
\alpha_{A}(r)=\frac{u_{A}\left(  r\right)  }{u_{A}^{\prime}\left(  r\right)
}=\frac{I_{A}(r)-e^{2i\delta_{A}}I_{A}^{\ast}(r)}{I_{A}^{\prime}%
(r)-e^{2i\delta_{A}}I_{A}^{\prime\ast}(r)}. \label{alpha}%
\end{equation}
Then%
\begin{equation}
e^{2i\delta_{A}}=\frac{I_{A}(r)-\alpha_{A}(r)I_{A}^{\prime}(r)}{I_{A}^{\ast
}(r)-\alpha_{A}(r)I_{A}^{\prime\ast}(r)}, \label{exp2id_1}%
\end{equation}%
\begin{equation}
e^{2i\delta_{A}}\left[  I_{A}^{\ast}(r)-\alpha_{A}(r)I_{A}^{\prime\ast
}(r)\right]  =I_{A}(r)-\alpha_{A}(r)I_{A}^{\prime}(r). \label{exp2id_2}%
\end{equation}
We place a dot on top of a function to indicate the derivative with respect to
$p_{A}$. \ Differentiating Eq.~(\ref{exp2id_2}) with respect to $p_{A}$, we
get%
\begin{align}
&  2ie^{2i\delta_{A}}\dot{\delta}_{A}\left[  I_{A}^{\ast}(r)-\alpha
_{A}(r)I_{A}^{\prime\ast}(r)\right]  +e^{2i\delta_{A}}\left[  \dot{I}%
_{A}^{\ast}(r)-\alpha_{A}(r)\dot{I}_{A}^{\prime\ast}(r)-\dot{\alpha}%
_{A}(r)I_{A}^{\prime\ast}(r)\right] \nonumber\\
&  =\dot{I}_{A}(r)-\alpha_{A}(r)\dot{I}_{A}^{\prime}(r)-\dot{\alpha}%
_{A}(r)I_{A}^{\prime}(r). \label{exp2id_ddot}%
\end{align}
\ Solving for $\dot{\delta}_{A}$ gives%
\begin{equation}
\dot{\delta}_{A}=F_{A}(r)+G_{A}(r)\dot{\alpha}_{A}(r), \label{F_G}%
\end{equation}
where%
\begin{equation}
F_{A}(r)=-\frac{1}{2i}\frac{\dot{I}_{A}^{\ast}(r)-\alpha_{A}(r)\dot{I}%
_{A}^{\prime\ast}(r)-e^{-2i\delta_{A}}\left[  \dot{I}_{A}(r)-\alpha_{A}%
(r)\dot{I}_{A}^{\prime}(r)\right]  }{I_{A}^{\ast}(r)-\alpha_{A}(r)I_{A}%
^{\prime\ast}(r)} \label{F}%
\end{equation}
and%
\begin{equation}
G_{A}(r)=\frac{1}{2i}\frac{I_{A}(r)I_{A}^{\prime\ast}(r)-I_{A}^{\ast}%
(r)I_{A}^{\prime}(r)}{\left\vert I_{A}(r)-\alpha_{A}(r)I_{A}^{\prime
}(r)\right\vert ^{2}}. \label{G}%
\end{equation}
Both $F_{A}$ and $G_{A}$ can be simplified further. We replace $\alpha_{A}$
using Eq.~(\ref{alpha}) and find%
\begin{equation}
F_{A}(r)=-\frac{1}{i}\frac{\operatorname{Re}\left\{  \dot{I}_{A}(r)\left[
I_{A}^{\prime\ast}(r)-e^{-2i\delta_{A}}I_{A}^{\prime}(r)\right]  -\dot{I}%
_{A}^{\prime}(r)\left[  I_{A}^{\ast}(r)-e^{-2i\delta_{A}}I_{A}(r)\right]
\right\}  }{I_{A}^{\ast}(r)I_{A}^{\prime}(r)-I_{A}(r)I_{A}^{\prime\ast}(r)}.
\end{equation}
The Wronskian identity, Eq.~(\ref{I_Wronskian}), leads to%
\begin{equation}
F_{A}(r)=-\frac{1}{2p_{A}^{2L+d-2}}\operatorname{Re}\left\{  \dot{I}%
_{A}(r)I_{A}^{\prime\ast}(r)-\dot{I}_{A}^{\prime}(r)I_{A}^{\ast}%
(r)-e^{-2i\delta_{A}}\left[  \dot{I}_{A}(r)I_{A}^{\prime}(r)-\dot{I}%
_{A}^{\prime}(r)I_{A}(r)\right]  \right\}  .
\end{equation}
With the same Wronskian identity, $G_{A}$\ simplifies to%
\begin{equation}
G_{A}(r)=\frac{p_{A}^{2L+d-2}}{\left\vert I_{A}(r)-\alpha_{A}(r)I_{A}^{\prime
}(r)\right\vert ^{2}}.
\end{equation}
We note that%
\begin{equation}
\dot{\alpha}_{A}(r)=\lim_{p_{B}\rightarrow p_{A}}\frac{\frac{u_{B}\left(
r\right)  }{u_{B}^{\prime}\left(  r\right)  }-\frac{u_{A}\left(  r\right)
}{u_{A}^{\prime}\left(  r\right)  }}{p_{B}-p_{A}}=\lim_{p_{B}\rightarrow
p_{A}}\frac{W[u_{B},u_{A}](r)}{\left(  p_{B}-p_{A}\right)  u_{B}^{\prime
}(r)u_{A}^{\prime}(r)}.
\end{equation}
For any $p_{A}\neq0$, we use Eq.~(\ref{nonthreshold_integral}) to get%
\begin{equation}
\dot{\alpha}_{A}(r)=\frac{2p_{A}}{\left[  u_{A}^{\prime}\left(  r\right)
\right]  ^{2}}\int_{0}^{r}dr^{\prime}\,\left[  u_{A}\left(  r^{\prime}\right)
\right]  ^{2}.
\end{equation}
We see that both $G_{A}(r)$ and $\dot{\alpha}_{A}(r)$ are non-negative and so%
\begin{equation}
\dot{\delta}_{A}=F_{A}(r)+G_{A}(r)\dot{\alpha}_{A}(r)\geq F_{A}(r).
\end{equation}
This inequality holds for all $p_{A}\neq0,$ and therefore also holds in the
limit $p_{A}\rightarrow0$. \ This is Wigner's causality bound generalized to
arbitrary dimension $d$.

Away from threshold, the equivalence between Wigner's bound and
Eq.~(\ref{nonthreshold_causality_bound}) is clear from the biconditional
statement,%
\begin{equation}
\dot{\delta}_{A}\geq F_{A}(r)\Leftrightarrow\dot{\alpha}_{A}(r)=\lim
_{p_{B}\rightarrow p_{A}}\frac{W[u_{B},u_{A}](r)}{\left(  p_{B}-p_{A}\right)
u_{B}^{\prime}(r)u_{A}^{\prime}(r)}\geq0.
\end{equation}
For $p_{A}=0$, the equivalence with our causality bounds follows from the
low-energy expansion for $W[u_{B},u_{A}]$,%
\begin{align}
W[u_{B},u_{A}](r)  &  =p_{B}^{2}\left\{  \frac{1}{2}r_{L,d}W[s_{0}%
,c_{0}](r)+\left(  \frac{1}{a_{L,d}}\right)  ^{2}W[s_{2},s_{0}](r)\right.
\nonumber\\
&  \left.  -\frac{1}{a_{L,d}}W[s_{2},c_{0}](r)-\frac{1}{a_{L,d}}W[c_{2}%
,s_{0}](r)+W[c_{2},c_{0}](r)\right\}  +O(p_{B}^{4})\text{.}%
\end{align}
\bibliographystyle{apsrev}
\bibliography{WignerPaper}

\end{document}